\documentclass[smallextended]{svjour3}
\usepackage{xcolor,graphicx,import,fullpage,textcomp,colortbl,array,pgfplots,lscape,todonotes,booktabs,dsfont,marvosym,ulem} \usepackage[fleqn]{amsmath}
\usepackage{graphicx,subcaption,caption,pgfplots,amsfonts}
	\newlength\figureheight
	\newlength\figurewidth
\definecolor{grey}{gray}{0.5}
\definecolor{lightgrey}{gray}{0.8}
\usepackage[sort,numbers]{natbib} 

\usepackage{nccmath}

\newcommand{\ra}[1]{\renewcommand{\arraystretch}{#1}}
\usepackage{multirow}
\usepackage[colorlinks=true, linkcolor=black, citecolor=black, urlcolor=black]{hyperref}
\usepackage{glossaries,textgreek,bm}
\usepackage[framed,numbered,autolinebreaks,useliterate]{mcode}
\usepackage[toc,page]{appendix}
\newglossary[slg]{symbolslist}{syi}{syg}{List of symbols} 
\makeglossaries 
\usepackage{glossaryEntries}
\usepackage{amsmath}
\usepackage{algorithm,algpseudocode}
\numberwithin{equation}{section}
\definecolor{darkred}{rgb}{.8,0,0}
\definecolor{darkgreen}{rgb}{0,.6,0}

\newcommand{\ca}{Ca$^{\text{\scriptsize 2+}}$}

\newcommand{\na}{Na$^{\text{\scriptsize +}}$}
\newcommand{\pot}{K$^{\text{\scriptsize +}}$}

\newacronym{smc}{SMC}{smooth muscle cell}
\newacronym{ec}{EC}{endothelial cell}
\newacronym{cicr}{CICR}{Ca$^{2+}$ induced Ca$^{2+}$ release}
\newacronym{nvc}{NVC}{neurovascular coupling}
\newacronym{ip3}{IP$_3$}{inotisol trisphosphate}
\newacronym{nvu}{NVU}{neurovascular unit}
\newacronym{sr}{SR}{sarcoplasmic reticulum}
\newacronym{er}{ER}{endoplasmic reticulum}
\newacronym{hopf}{H}{Hopf}
\newacronym{fp}{FP}{fixed point}
\newacronym{ode}{ODE}{ordinary differential equation}
\newacronym{pde}{PDE}{partial differential equation}
\newacronym{lpc}{LPC}{limit point cycle}
\newacronym{lp}{LP}{limit point}
\newacronym{kir}{KIR}{inward rectifying K$^+$}
\newacronym{pvs}{PVS}{perivascular space}
\newacronym{ne}{NE}{neuron}
\newacronym{sc}{SC}{synaptic cleft}
\newacronym{ac}{AC}{astrocyte}
\newacronym{bt}{BT}{Bogdanov-Takens}
\newacronym{cp}{CP}{Cusp}
\newacronym{gh}{GH}{Generalised Hopf}
\newacronym{ic}{IC}{initial condition}
\newacronym{bc}{BC}{boundary condition}
\newacronym{csd}{CSD}{cortical spreading depression}
\newacronym{mpi}{MPI}{Message Passing Interface}
\newacronym{vtk}{VTK}{Visualisation Toolkit}
\newacronym{ghk}{GHK}{Goldman Hodgkin Katz}
\newacronym{plc}{PLC}{phospholipase-C}
\newacronym{pd}{PD}{Period Doubling}
\newacronym{fhn}{FHN}{FitzHugh-Nagumo}
\newacronym{2d}{2D}{two dimensional}
\newacronym{bk}{BK}{big potassium}
\newacronym{vocc}{VOCC}{voltage operated \gls{ca} channel}
\newacronym{cbf}{CBF}{cerebral blood flow}
\newacronym{ca}{Ca$^{2+}$}{calcium}
\newacronym{pot}{K$^+$}{potassium}
\newacronym{cl}{Cl$^1$}{chlorine}
\newacronym{na}{Na$^+$}{sodium}
\newacronym{atp}{ATP}{adenosine triphosphate}
\newacronym{rhs}{RHS}{right hand side}

\newacronym{trpv}{TRPV4}{transient receptor potential vanniloid-related 4}
\newacronym{no}{NO}{nitric oxide}
\newacronym{20hete}{20-HETE}{20- hydroxyeicosatetraenoic acid}
\newacronym{eet}{EET}{epoxyeicosatrienoic acid}
\newacronym{cox}{COX}{cyclooxegenase enzymes}
\newacronym{aa}{AA}{arachidonic acid}
\newacronym{PgE2}{PgE$_2$}{prostaglandin E$_2$}
\newacronym{ecs}{ECS}{extracellular space}
\newacronym{wss}{WSS}{wall sheer stress}
\newacronym{nmda}{NMDA}{N-methyl-D-aspartate}
\newacronym{sgc}{sGC}{soluble guanylyl cyclase}
\newacronym{cgmp}{cGMP}{cyclic guanosine monophosphate}
\newacronym{mglur}{mGluR}{metabotropic glutamate receptor}

\newacronym{efs}{EFS}{electro field stimulation}
\newacronym{mlc}{MLC}{myosin light chain kinase}
\newacronym{pkc}{PKC}{protein-kinase C}
\newacronym{cpi17}{CPI-17}{myosin phosphatase inhibitor protein}

\newacronym{bold}{BOLD}{blood-oxygen-level dependent}
\newacronym{fmri}{fMRI}{functional magnetic resonance imaging}
\newacronym{cbv}{CBV}{cerebral blood volume}
\newacronym{nat}{NaT}{transient Na$^+$}

\newacronym{sodpot}{Na$^+$/K$^+$}{sodium potassium}
\newacronym{nnos}{nNOS}{neuronal \gls{no} synthase}
\newacronym{enos}{eNOS}{endothelial \gls{no} synthase}
\newacronym{cmro2}{CMRO$_2$}{cerebral metabolic rate of oxygen}
\newacronym{hbo}{HbO}{oxyhemoglobin}
\newacronym{hbr}{HbR}{deoxyhemoglobin}
\newacronym{hbt}{HbT}{total hemoglobin}
\newacronym{mtt}{MTT}{mean transit time}
\newacronym{ltp}{LTP}{long term potentiation}
\newacronym{epsp}{EPSP}{excitatory postsynaptic potential}
\newacronym{ipsp}{IPSP}{inhibitory postsynaptic potential}
\newacronym{lc}{LC}{locus coeruleus}
\newacronym{lcna}{LC-NA}{noradrenalin locus coeruleus}

\begin{document}
\title{Global Sensitivity Analysis of High Dimensional Neuroscience Models: An Example of Neurovascular Coupling}
\author{ J.L. Hart\thanks{J.L.H. was supported in part by the National Science Foundation (NSF) award NSF DMS-1522765.} \and P.A. Gremaud  \thanks{P.A.G. was supported in part by National Science Foundation (NSF) awards NSF DMS-1522765 and DMS-1745654.}\and T. David}
\institute{J.L. Hart \at Department of Mathematics, North Carolina State University, Raleigh, NC, USA, \email{jlhart3@ncsu.edu}, \and P.A. Gremaud \at Department of Mathematics, North Carolina State University, Raleigh, NC, USA, \email{gremaud@ncsu.edu} \and T. David \at Department of Mechanical Engineering, University of Canterbury, New Zealand, \email{tim.david@canterbury.ac.nz}}
\maketitle
\thispagestyle{empty}

\begin{abstract}
The complexity and size of state-of-the-art cell models have significantly increased in part due to  the requirement that these models possess complex cellular functions which are thought--but not necessarily proven--to be important. Modern cell models often involve hundreds of parameters; the values of these parameters come, more often than not, from animal experiments whose relationship to the human physiology is weak with very little information on the errors in these measurements. The concomitant uncertainties in parameter values result in uncertainties in the model outputs or Quantities of Interest (QoIs). Global Sensitivity Analysis (GSA) aims at apportioning to individual parameters (or sets of parameters) their relative contribution to output uncertainty thereby introducing a measure of influence or importance of said parameters. New GSA approaches are required to deal with increased model size and complexity; a three stage methodology consisting of screening (dimension reduction), surrogate modeling, and computing Sobol' indices, is presented. The methodology is used to analyze a physiologically validated numerical model of neurovascular coupling which possess 160 uncertain parameters. The sensitivity analysis investigates three quantities of interest (QoIs), the  average value of $K^+$ in the extracellular space, the average volumetric flow rate through the perfusing vessel, and the minimum value of the actin/myosin complex in the smooth muscle cell. GSA provides a measure of the influence of each parameter, for each of the three QoIs, giving insight into areas of possible physiological dysfunction and areas of further investigation. 

\end{abstract}

\section{Introduction}\label{sec:intro}
Over the past 20 years, the use of computational models to describe physiological phenomena has grown spectacularly. This growth has provided significant advantages to the experimental community in that it can furnish results of \textit{in silico} experiments which are either ethically or physically impossible in the laboratory. However, these models have also increased in complexity with a concomitant increase in the associated number of parameters (for a variety of reasons, most notably the requirement that the models should possess complex cellular functions which are thought \textit{but not necessarily proven} to be important). These parameters, in defining the relevant phenomenon, more often than not come from a plethora of animal experiments whose relationship to the human physiology is weak.  In addition, experimental results in the public domain provide very little information on the errors in these measurements. 

Even simple physiological models tend to be  highly non-linear; their range of applicability and reliability can only be assessed  through careful analysis. For large complex systems, the sensitivity of quantities of interest to model parameters is a priori unclear. Herein lies one of the difficulties of modeling: what effect do uncertainties in parameters determined from experiment have on the output of a non-linear numerical model? Because of both model complexity and high dimensionality, sensitivity analysis is  computationally demanding and may require several ad hoc steps--such as screening (reducing the parameter dimension). Importantly, in analysing  sensitivities, we can learn significant facts about the physiology of the system which would have stayed hidden under the premise of simply producing results. 
From a purely physiological perspective, an understanding of the dominant cellular mechanisms resulting in cerebral tissue perfusion after neuronal stimulation would be of particular and important interest. 

In the above context, we investigate the sensitivity of  the \gls{nvc} response (see Section~\ref{sec:model}), which has a large parameter dimension where most (if not all) of the parameter values come from non-human experiment  with an inherent (unknown) error. We denote by $\mathbf y = (y_1, \dots, y_N)$  the {\sl state variables} of the model and by $\boldsymbol{\theta} = (\theta_1, \dots, \theta_P)$ the {\sl uncertain parameters} of the model. The evolution of the state variables is governed by a system of ordinary differential equations (ODEs) 
\begin{eqnarray}
\frac {d\boldsymbol{y}}{dt} = \mathbf{f}(\mathbf{y}, \boldsymbol{\theta}), \label{caboodle}
\end{eqnarray}
where $\mathbf{f}$ is a known function of its arguments. Equation (\ref{caboodle}) is completed with a set of initial conditions $\boldsymbol{y}(0) = \boldsymbol{y}_0$;  here, we simply take $\boldsymbol{y}_0$ as the  equilibrium solution at the parameters' nominal values, i.e., $\mathbf f(  \mathbf{y}_0, \bar{\boldsymbol{\theta}}) = 0$, where $ \bar{\boldsymbol{\theta}} = (\bar\theta_1, \dots, \bar\theta_P)$ denotes the nominal values of the parameters. The Supplementary Material contains code and the nominal values of parameters, further information can be found in \cite{Dormanns2015a}.

Based on physiological considerations, we examine three quantities of interest (QoI), see Section~\ref{sec:model}.  Let $q$ be one of our three  QoIs; while determined from the state variables, i.e., from $\mathbf y$, $q$ is  ultimately a function of the parameters alone (and possibly time), i.e.
\begin{eqnarray}
q = g(\boldsymbol{\theta}). \label{qoi}
\end{eqnarray}
The overall goal of our numerical study is to determine  which of the uncertain parameters $\theta_1, \dots, \theta_P$  are the most/least influential for each QoI. There is a substantial amount of research currently being done in applied mathematics and statistics in the corresponding field of Global Sensitivity Analysis  (GSA). How to meaningfully define ``influential" or ``non-influential" and  how to develop methods applicable to high-dimensional problems are two significant challenges of the field \cite{corvar,timegsa,stogsa,iooss,owen,saltelli}.

The present \gls{nvc} model contains $N=67$ state variables and $P= 160$ uncertain parameters. The complexity of the model and large parameter space dimension preclude a direct application of GSA tools. Part of our contribution in this paper is to show how multiple GSA tools may be combined to analyse such problems.

\section{Physiological Model: Neurovascular Coupling}\label{sec:model}
The \gls{nvc} response, i.e., the ability to locally adjust vascular resistance as a function of neuronal activity, is believed to be mediated by a number of different signaling mechanisms. A  mechanism based on a metabolic negative feedback theory was first proposed in \cite{Roy1890}. According to this theory, neural activity leads to a drop in oxygen or glucose levels and increases in CO$_2$, adenosine, and lactate levels. All of these signals could dilate arterioles and hence were believed to be part of the neurovascular response. However, recent experiments illustrated that the \gls{nvc} response is partially independent of these metabolic signals \cite{Leithner2010, Lindauer2010, Mintun2001, Powers1996, Makani2010}. An alternative to this theory was proposed where the neuron releases signaling molecules to directly or indirectly affect the blood flow. Many mechanisms such as the \gls{pot} signaling mechanism \cite{Filosa2006}, the \gls{no} signaling mechanism, or the arachidonic acid to \gls{eet} pathway are found to contribute to the neurovascular response \cite{Attwell2010}.

The \gls{pot} signaling mechanism of \gls{nvc} seems to be supported by significant evidence, although new evidence shows that the endfoot astrocytic \gls{ca} could play a significant role. The \gls{pot} signaling hypothesis mainly utilises the astrocyte,  positioned to enable the communication between the neurons and the local perfusing blood vessels. The astrocyte and the \glspl{ec} surrounding the perfusing vessel lumen exhibit a striking similarity in ion channel expression and thus can enable control of the \gls{smc} from both the neuronal and blood vessel components \cite{Longden2015}. Whenever there is neuronal activation \gls{pot} ions are released into the \gls{ecs} and \gls{sc}. The astrocyte is depolarised by taking up \gls{pot} released by the neuron and releases it into the \gls{pvs} via the endfeet through the BK channels \citep{Filosa2007}. This increase in \gls{ecs} \gls{pot} concentration ($3-10$ mM) near the arteriole hyperpolarises the \gls{smc} through the \gls{kir} channel, effectively closing the voltage-gated \gls{ca} channel, reducing smooth muscle cytosolic \gls{ca} and thereby causing dilation. Higher \gls{pot} concentrations in the \gls{pvs} cause contraction due to the reverse flux of the \gls{kir} channel \citep{Farr2011}.

In maintaining a relatively homeostatic condition the neuron uses a considerable amount of energy ( as noted below) on ensuring a specific concentration of ions (and ion gradients) in the extracellular space. This concentration of \pot and \na can be envisaged as one of the inputs to the astrocyte in determining the nutrient flux to the cerebral tissue. Parameters which model/affect the energy usage are therefore important in simulating neurovascular coupling. In order not to increase the complexity of the numerical model further than is necessary an approximation needs to be made on how to model the energy usage and the associated parameters. Estimates of the relative demands of the cerebral processes that require energy were given based on different experimental data by \citet{Ames2000}. Vegetative processes that maintain this homeostasis including protein synthesis accounted for $10-15$\% of the total energy consumption. The costliest function seems to be in restoring the ionic gradients during neural activation. The \gls{sodpot} exchange pump is estimated to consume $40-50$\%, while the \gls{ca} influx from organelles and extracellular fluid consumes $3-7$\%. Processing of neurotransmitters such as uptake or synthesis consumes $10-20$\%, while the intracellular signaling systems which includes activation and inactivation of proteins consumes $20-30$\%. The rest of the energy is estimated to be consumed by the axonal and dendritic transport in both directions. Given the distribution of energy amongst the cellular pathways it is assumed for this model that variation in oxygen concentration in the cerebral tissue affects only the \gls{sodpot} exchange pump. 

\begin{figure}[h!]
\centering
\includegraphics[width=0.7\linewidth]{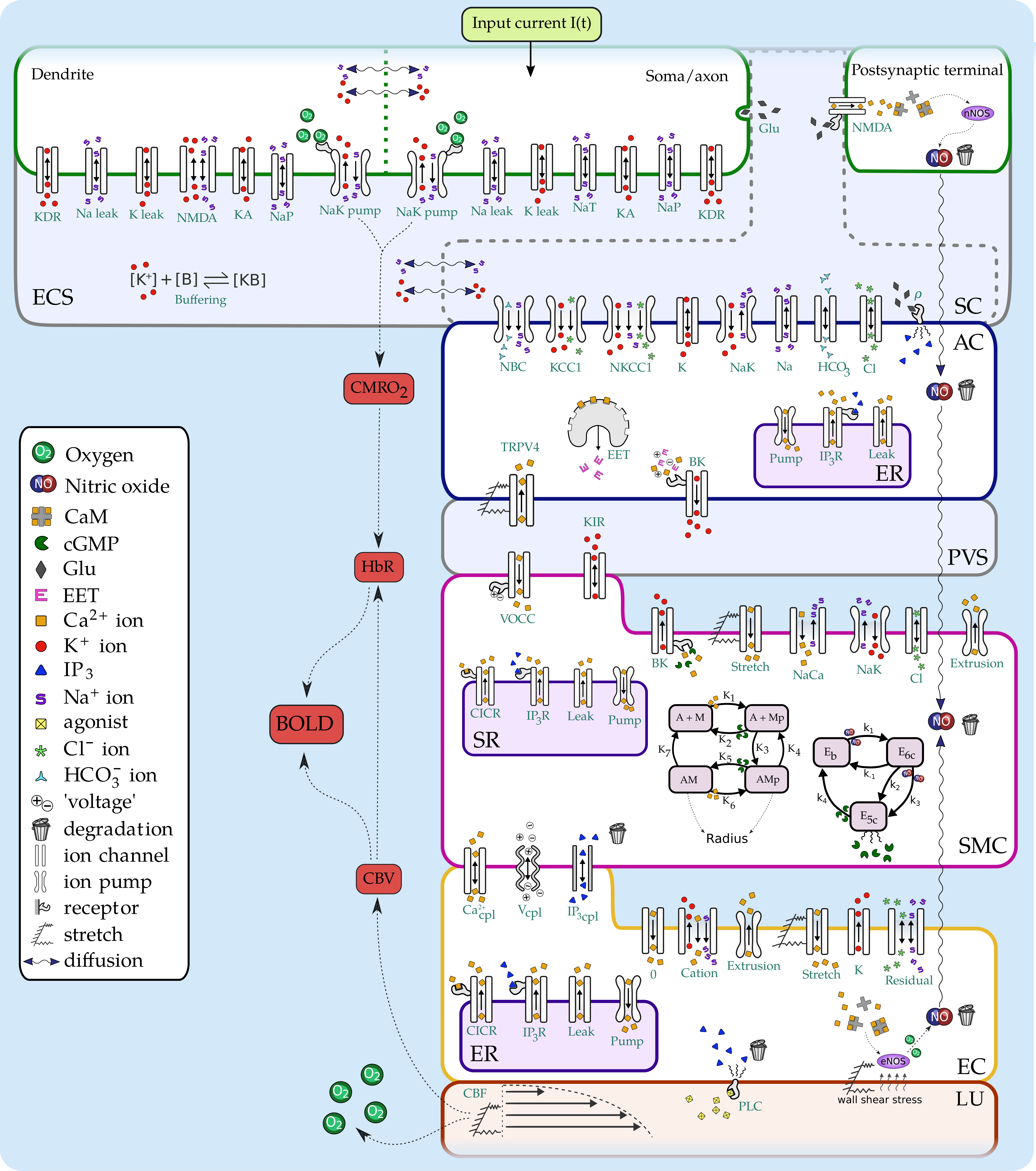}
\caption[Graphic Sketch of NVU version 2.0]{Graphic Sketch of NVU version 2.0, showing the basic components of neuron, astrocyte, smooth muscle cell, endothelial cell, lumen and extracellular space. ion channels, pumps and pathways from neuron to endothelial cell are also shown in addition to the basis for evaluation of the fMRI blood oxygenation level signal (BOLD) protocol.}
\label{fig:nvu20}
\end{figure}

Previous work \cite{Mathias2018} has provided  the construction of an experimentally validated numerical (\textit{in silico}) model based on experimental data to simulate the \gls{fmri} \gls{bold} signal associated with \gls{nvc} along with the associated metabolic and blood volume responses. An existing neuron model \citep{Mathias2017, Mathias2017a} has been extended to include an additional transient \gls{na} ion channel (NaT) expressed in the neuron, and integrated into a complex \gls{nvc} model \citep{Dormanns2015, Dormanns2016b, Kenny2017a}. This present model is based on the hypothesis that the \gls{pot} signaling mechanism of \gls{nvc} is the primary contributor to the vascular response and the \gls{sodpot} exchange pump in the neuron is the primary consumer of oxygen during neural activation. The model contains 317 parameters, most of which come from non-human experiments. Based on the work by \cite{Dormanns2016b} and \cite{Kenny2018}, we have chosen a subset of parameters defining basic pathways, such as  the nitric oxide and potassium pathways,   that are considered important for the normal function of neurovascular coupling.  We model the uncertainty of the chosen parameters  by representing them as random variables. The remaining  parameters are fixed to nominal values; they include leak terms, characteristic oxygen and other species concentrations, buffer concentrations, volume surface ratios etc\dots By permitting variability in only the parameters that support these pathways, the dimension of the parameter space is reduced from 317 to 160, which greatly facilitates our analysis. The algorithms defined below can be used to investigate other complex models including that of neurovascular coupling. However, for this initial work, we constrain ourselves to the above subset. 
Figure \ref{fig:nvu20} shows the components and main pathways of the neurovascular coupling model (version 2.0). The numerical model outlined in sketch form in Figure \ref{fig:nvu20} is fully defined in the Supplementary Material  and has been developed over a number of years \cite{Farr2011,Dormanns2015,Dormanns2016b}.

\begin{figure}[h]
\centering
\includegraphics[width=.4 \textwidth]{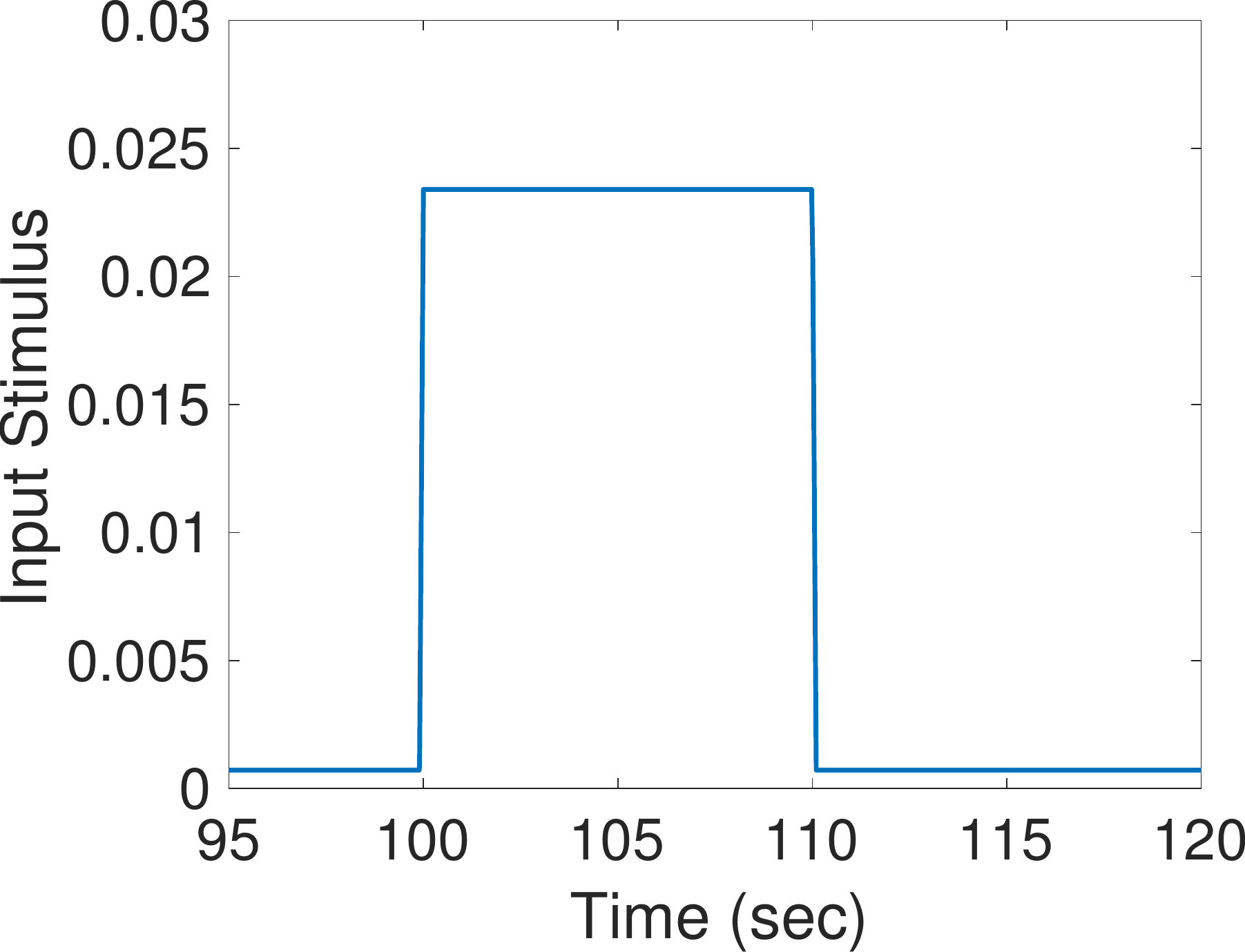}
\includegraphics[width=.4 \textwidth]{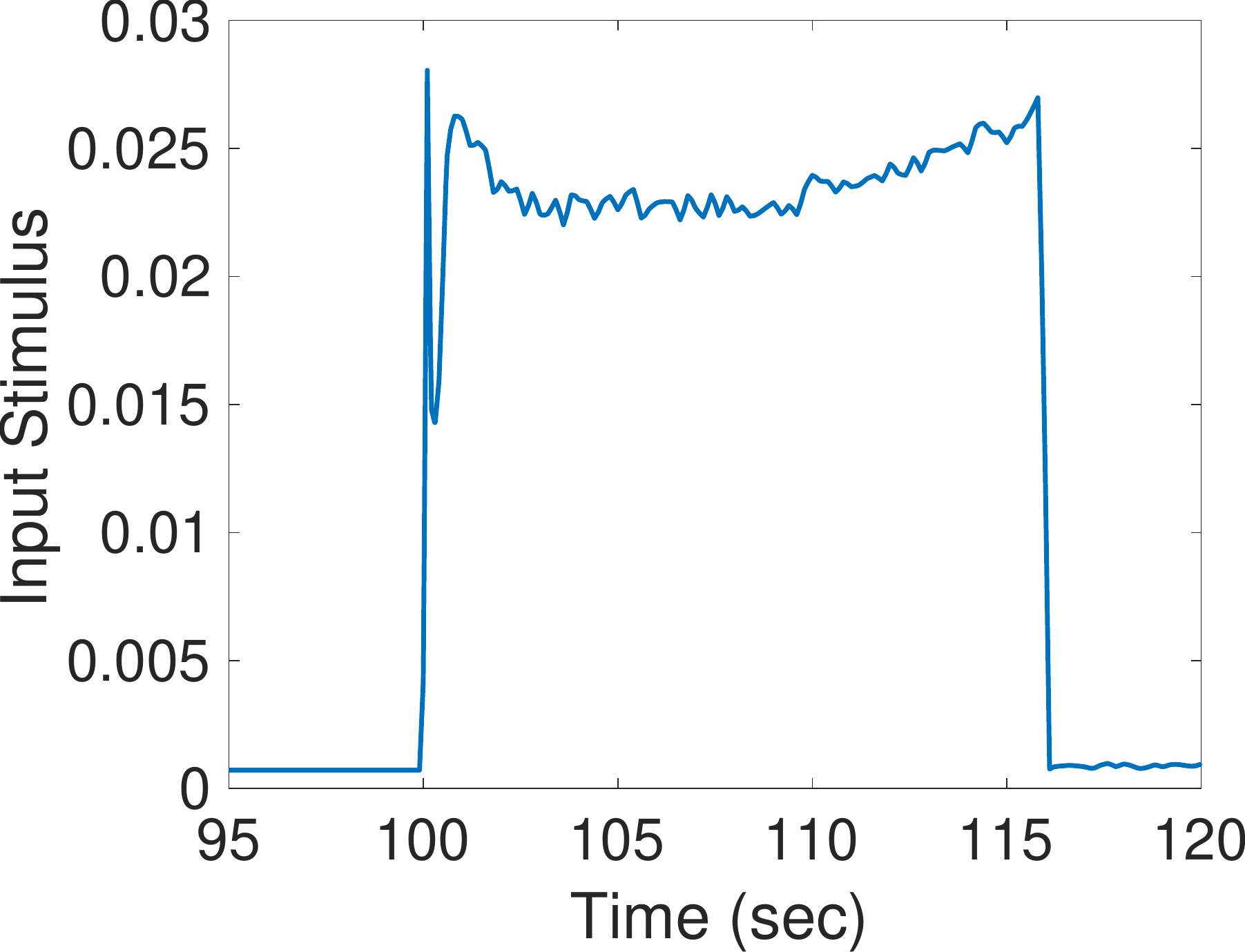}
\caption{Left: rectangular pulse input stimulus; right: stimulus used in lab experiments.}
\label{input_stimuli}
\end{figure}

In order to induce a variation in radius of a vessel which perfuses the associated cerebral tissue following a neuronal stimulation, an input current is used. The numerical experiments solve equation (\ref{caboodle}) in the presence of short duration electrical (current) stimuli as displayed in Figure~\ref{input_stimuli}. For the presented cases, two input profiles are utilised. A rectangular pulse of width 10 seconds and a experimental pulse sequence (stimulating the whisker pad of a rat) used in the work of \cite{Zheng2010} which has the same magnitude as the rectangular pulse but a duration of sixteen seconds followed by second pulse (which is not used in this analysis). Further information about the experiment and the results can be found in \cite{Zheng2010}.  For this second case, the stimulus was a current injection at the rat whisker pad. This induced an increase in neuronal activity in the somato-sensory cortex which subsequently, through the neurovascular pathway, produced a  change in the radius $R(t)$ allowing increased nutrients to perfuse into the cerebral tissue; this is  the essence of neuro-vascular coupling.

 We wish to analyse parameter sensitivity of quantities of interest (QoIs) that are key to our understanding and quantification of  neurovascular coupling.   From previous work, it is known that the \pot concentration in the extracellular space (ECS) is crucial to maintaining a homeostatic state (large concentrations of \pot support the propagation of spreading depression waves) hence we choose to look at the average value of \pot in the ECS as our first QoI $q_1$. Secondly, neurovascular coupling is the main phenomenon providing oxygen and nutrients to the neuronal tissue. This is mediated by the local arteriole dilating and (under the assumption of constant pressure) increasing the flow of blood into the tissue. We therefore define the time averaged cerebral blood flow determined over the course of neuronal stimulation as our second QoI, $q_2$. The contraction/dilation of the arteriolar vessel  depends on the smooth muscle cell (SMC) concentration of \ca and the phosphorylation of the actin/myocin complex. Our third QoI $q_3$ is the minimum combined concentration of the actin myosin complex, both phosphorylated and unphosphorylated. This allows us to analyse the functioning of the main components of the neurovascular unit (NVU) defined as the linked components of neuron, synaptic ceft, astrocyte, perivascular space (PVS), SMC, endothelial cell (EC), lumen (LU, the domain in which blood flows) and the ECS. 
  Assuming the stimulation occurs for time $t$ between $t_1$ and $t_2$, $t_2>t_1$,  the QoIs are thus
 \begin{itemize}
\item average ECS potassium 
\begin{eqnarray}
 q_1 = \frac{1}{t_2-t_1}\int_{t_1}^{t_2}[K^+]_{ECS}(s)\, ds, \label{K_ECS_Mean}
\end{eqnarray}
\item average volumetric flow rate in the cerebral tissue
\begin{eqnarray}
 q_2 = \frac{1}{t_2-t_1}\int_{t_1}^{t_2}\left(\frac{R(s)}{R_0}\right)^4\, ds, \label{vol_flow}
\end{eqnarray}
 \item minimum combined concentration of the actin myosin complex, both phosphorylated and unphosphorylated
\begin{eqnarray}
q_3 = \min _{t, t_1\le t \le t_2} [AM(t)+AM_p(t)], \label{AM_AMp_Min}
\end{eqnarray}
which we will refer to as $[AM+AM_p]_{min}$ throughout the article.
\end{itemize}

Other examples of sensitivity analysis studies relevant to the general field of biomedicine include \cite{gsa_pharm,lr_gsa,uqpy,Witthoft2013}.
  
\section{Methodology}\label{sec:meth}

Our analysis may be described in a four step process:
\begin{enumerate}
\item[(i)] define a probability distribution for $\theta$ which models its uncertainty,
\item[(ii)] draw samples from this distribution, 
\item[(iii)] evaluate the QoI for each of these parameter samples,
\item[(iv)] use these QoI evaluations to infer its sensitivity to each parameter. 
\end{enumerate}
Subsection~\ref{sec:param_dist_fit} describes our approach for step (i). Step (ii) is easily executed using standard methods. The computational bottleneck for our analysis is the QoI evaluations (which require ODE solves) in step (iii). While relatively large and stiff, the ODE system (\ref{caboodle}) can be  solved with standard tools and methods, here through the MATLAB routine ode15s with relative and absolute tolerances of $10^{-4}$. The evaluation of the above QoIs themselves from the ODE solutions is straightforward and can be done at low cost. Step (iv), inferring the sensitivities, may be done in a plurality of ways. Because of the parameter dimension and computational cost of model evaluations, we perform step (iv) with a multi-phased procedure,
\begin{enumerate}
\item[(I)] screening,
\item[(II)] surrogate modeling,
 \item[(III)] and computing Sobol' indices,
 \end{enumerate}
 described in Subsections~\ref{sec:screen}, ~\ref{sec:surrogate}, and ~\ref{sec:gsa} respectively. We summarize our method in Subsection~\ref{sec:summary}.

\subsection{Parameter Distribution Fitting}
\label{sec:param_dist_fit}

Describing the uncertainty attached to the parameter vector $\boldsymbol{\theta}$ is an important and delicate {\sl modeling assumption}. Here, we give each $\theta_i$, $i=1,\dots, 160$, a nominal value $\bar \theta_i$ and assume each parameter to be independent and uniformly distributed over the interval $[0.9\, \bar\theta_i, 1.1 \,\bar\theta_i]$, i.e., within $\pm 10\%$ of the nominal value. Larger intervals were considered for the parameters. For the NVU model under consideration, increasing the width of the interval resulted in many samples for which the solution exhibited atypical or non-physical behavior, or in some cases ode15s was unable to solve the system. Our choice of $\pm 10\%$ uncertainty is considered a reasonable compromise between accounting for uncertainty and ensuring computational feasibility. 
	
Our initial assumption of parameter independence is incorrect. In fact, ode15s is unable to solve the system for most samples drawn under this assumption. The parameter dependencies, which are unknown a priori, are discovered through a computational procedure akin to Approximate Bayesian Computation \cite{abc}. Our approach, described below, is computationally intractable when applied to the NVU model under consideration. We simplify the model by removing the stimulus, i.e. modeling the steady state behavior of the system. A collection of $S$ samples, $\boldsymbol{\theta}^k$, $k=1,2,\dots,S$, are drawn from our initial distribution (assuming independence) and the ODE system is solved for each parameter sample. The collection of solutions are post-processed to partition the samples $\{\boldsymbol{\theta}^k\}_{k=1}^S$ into two subsets $\{\boldsymbol{\theta}^{a_k}\}_{k=1}^{S_a}$ and $\{\boldsymbol{\theta}^{r_k}\}_{k=1}^{S_r}$, where $a_k$, $k=1,2,\dots,S_a$ denotes the samples where the steady state solution exhibited physiologically normal behavior and $r_k$, $k=1,2,\dots,S_r$ denotes the samples where it does not. We reject the samples $\{\boldsymbol{\theta}^{r_k}\}_{k=1}^{S_r}$ and fit a distribution to the accepted samples $\{\boldsymbol{\theta}^{a_k}\}_{k=1}^{S_a}$ using standard statistical methods. This new distribution is sampled $S$ times,  the ODE system is solved for each sample, and the results are post-processed into accepted and rejected samples again. We continue this process iteratively until satisfactory convergence of the fitted distribution.

\subsection{Screening}
\label{sec:screen}
Having determined a parameter distribution, we evaluate the QoI $q=g(\boldsymbol{\theta})$ at $M$ different samples, denote them $\boldsymbol{\theta}^k$, $k=1,2,\dots,M$.
We fit a {\sl linear model} to the QoI under study
\begin{eqnarray}
g(\boldsymbol\theta^k) = g(\theta_1^k, \dots, \theta_P^k) \approx \beta_0 + \sum\limits_{j=1}^{160} \beta_j \theta_j^k, \quad k=1, \dots, M. \label{lr}
\end{eqnarray}
This approach yields a crude (but highly efficient here) sensitivity analysis of the model with respect to the $\theta_j$'s, $j=1,\dots, 160$. We assign a preliminary importance measure to each $\theta_j$ by computing for each of them the relative size of their coefficient in the above linear approximation, i.e.
\begin{eqnarray*}
L_j = \frac{\vert \beta_j \vert}{\sum\limits_{\ell=1}^{160} \vert \beta_\ell \vert}, \qquad j=1,\dots,160.
\end{eqnarray*}
To obtain a model with a more manageable size, we reduce the parameter space to only the $\theta_j$'s for which $L_j>0.01$. We denote these $r$ parameters $\{ \theta_{j_i}\}_{i=1}^r$. The rest of the parameters are regarded as non-influential and treated as latent variables, even though they are uncertain, their specific values (within the given range) have little bearing of the considered QoI. In other words, we consider the approximation 
\begin{eqnarray}
g(\theta_1, \dots, \theta_{160}) \approx h(\theta_{j_1}, \dots, \theta_{j_r}), \label{reddim}
\end{eqnarray}
where $h$ is obtained from $g$ by treating the non-influential parameters as latent. In Section~\ref{sec:results}, this reduction yields around 15-20 parameters instead of the original 160. We use $\hat{\boldsymbol{\theta}}$ to denote the reduced parameter vector.

\subsection{Surrogate model}
\label{sec:surrogate}
For any of the three considered QoIs, our information on the function $h$ defined (\ref{reddim}) consists of the set of sampled values $\{ h(\theta_{j_1}^k, \dots, \theta_{j_r}^k)\}$, $k=1, \dots, M$.  To facilitate the use of standard GSA tools, which may require derivatives or variance estimations, it is both convenient and computationally advantageous to construct an approximating function, i.e., a surrogate model. 
We use a sparse Polynomial Chaos (PC) surrogate. This amounts to introducing a polynomial approximation of $h$ of the type
\begin{eqnarray}
h(\hat{\boldsymbol{\theta}}) \approx H(\hat{\boldsymbol{\theta}}) \equiv \sum_{\boldsymbol{\alpha}} c_{\boldsymbol{\alpha}} \psi_{\boldsymbol{\alpha}}(\hat{\boldsymbol{\theta}}) \label{pce}
\end{eqnarray}
where the $\psi_{\boldsymbol{\alpha}}$'s are multivariate polynomials which are orthogonal with respect to the probability distribution function (PDF) $p_{\hat{\boldsymbol{\theta}}}$ of $\hat{\boldsymbol{\theta}}$, i.e.
\begin{eqnarray}
\int \psi_{\boldsymbol{\alpha}}(\mathbf x) \psi_{\boldsymbol{\beta}}(\mathbf x)\, p_{\hat{\boldsymbol{\theta}}}(\mathbf x) \, d\mathbf{x} = \delta _{\boldsymbol{\alpha},\boldsymbol{\beta}} \label{ortho}
\end{eqnarray}
where $\boldsymbol{\alpha}$ and $\boldsymbol{\beta}$ are multi-indices and $\delta _{\boldsymbol{\alpha},\boldsymbol{\beta}}$ is the generalized Kronecker symbol. The coefficients are computed through least-squares minimization, see the Appendix for additional discussion. All surrogate models are validated using 10-fold cross validation.

Polynomial Chaos is by now a well documented method. We use the \textit{UQLab} implementation for the results below and refer the reader to its manual \cite{uqlab} for more details.

\subsection{Sobol' indices} 
\label{sec:gsa}
We use variance based GSA to assess the relative importance of the input parameters of $H$ in (\ref{pce}). In their simplest form, the total Sobol' indices \cite{saltellitotalindex} apportion to uncertain parameters, or sets thereof, their relative contribution to the variance of the output. Indeed, thanks to the law of total variance, we can decompose the variance of $H(\hat{\boldsymbol{\theta}})$ as
\begin{eqnarray}
\operatorname{var}(H(\hat{\boldsymbol{\theta}})) = \operatorname{var}(\mathbb E[H(\hat{\boldsymbol{\theta}})|\hat{\boldsymbol{\theta}}_{\sim i}]) + \mathbb E[\operatorname{var}(H(\hat{\boldsymbol{\theta}})|\hat{\boldsymbol{\theta}}_{\sim i})], \label{ltv}
\end{eqnarray}
where $\hat{\boldsymbol{\theta}}_{\sim i}$ denotes all the parameters in $\hat{\boldsymbol{\theta}}$ except $\hat{\boldsymbol{\theta}}_i$. If we assume now that all the input parameters of $H$ are known with certainty, i.e., if $\hat{\boldsymbol{\theta}}_{\sim i}$ is known, then the remaining variance of $H(\hat{\boldsymbol{\theta}})$ is simply given by 
\begin{eqnarray*}
\operatorname{var}(H(\hat{\boldsymbol{\theta}})) - \operatorname{var}(\mathbb E[H(\hat{\boldsymbol{\theta}})|\hat{\boldsymbol{\theta}}_{\sim i}]) = \mathbb E[\operatorname{var}(H(\hat{\boldsymbol{\theta}})|\hat{\boldsymbol{\theta}}_{\sim i})]. 
\end{eqnarray*}
This latter expression is thus a natural way of measuring how influential $\hat{\boldsymbol{\theta}}_i$ is; the corresponding total Sobol' index $S_{T_i}$ is but a normalized version of this
\begin{eqnarray}
S_{T_i} = \frac{\mathbb E[\operatorname{var}(H(\hat{\boldsymbol{\theta}})|\hat{\boldsymbol{\theta}}_{\sim i})]}{\operatorname{var}(H(\hat{\boldsymbol{\theta}})) }. \label{sobol}
\end{eqnarray}
From this definition, one easily observes that $S_{T_i} \in [0,1]$; large values indicate that $\hat{\boldsymbol{\theta}}_i$ is important, $S_{T_i} \approx 0$ implies that $\hat{\boldsymbol{\theta}}_i$ is not important.
The relevance of this basic definition can be extended to time dependent QoIs \cite{timegsa}, stochastic QoIs \cite{stogsa} or correlated parameters \cite{corvar}.

\subsection{Summary of the Method}
\label{sec:summary}
Algorithm~\ref{algo} provides a summary of the method. 

\begin{algorithm}
\caption{overall numerical approach}\label{algo}
\begin{algorithmic}[1]
\While{parameter distribution has not converged}
\For{$k=1:S$} \Comment{sampling}
\State sample parameter distribution  $\longrightarrow \boldsymbol{\theta}^k$; solve (\ref{caboodle}) (without stimulus) $\longrightarrow \mathbf y^k$
\EndFor
\State partition the parameter samples into accepted and rejected samples
\State fit a new distribution to the accepted samples \Comment{see \S~\ref{sec:param_dist_fit}}
\EndWhile
\For{$k=1:M$} \Comment{sampling (final distribution)}
\State sample parameter distribution  $\longrightarrow \boldsymbol{\theta}^k$; solve (\ref{caboodle}) (with stimulus) $\longrightarrow \mathbf y^k$
\EndFor
\For{each QoI $q$}
\State solve the least-squares problem (\ref{lr}) \Comment{linear model}
\State identify the influential parameter vector $\hat{\boldsymbol\theta} =   (\theta_{j_1}, \dots, \theta_{j_r})$  \Comment{screening}
\State fit the polynomial chaos surrogate $H$ \eqref{pce} \Comment{surrogate model}
\State compute total Sobol' indices (\ref{sobol}) of the surrogate model $H$ \Comment{Sobol' indices}
\EndFor
\end{algorithmic}
\end{algorithm}

\section{Numerical results}
\label{sec:results}

\subsection{Parameter distribution}
\label{sec:param_sampling}

\begin{figure}[h]
\centering
\includegraphics[width=.4 \textwidth]{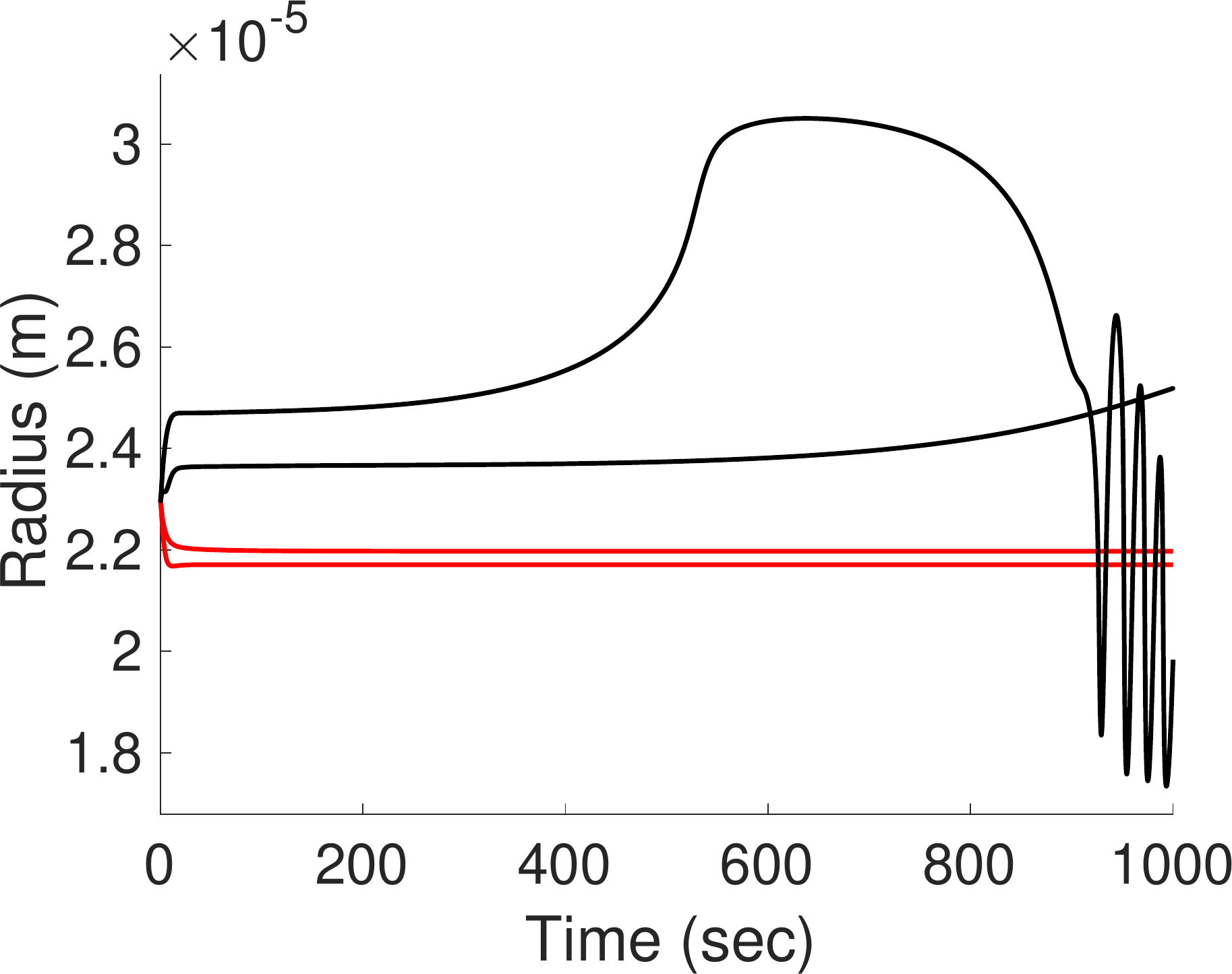}
\includegraphics[width=.4 \textwidth]{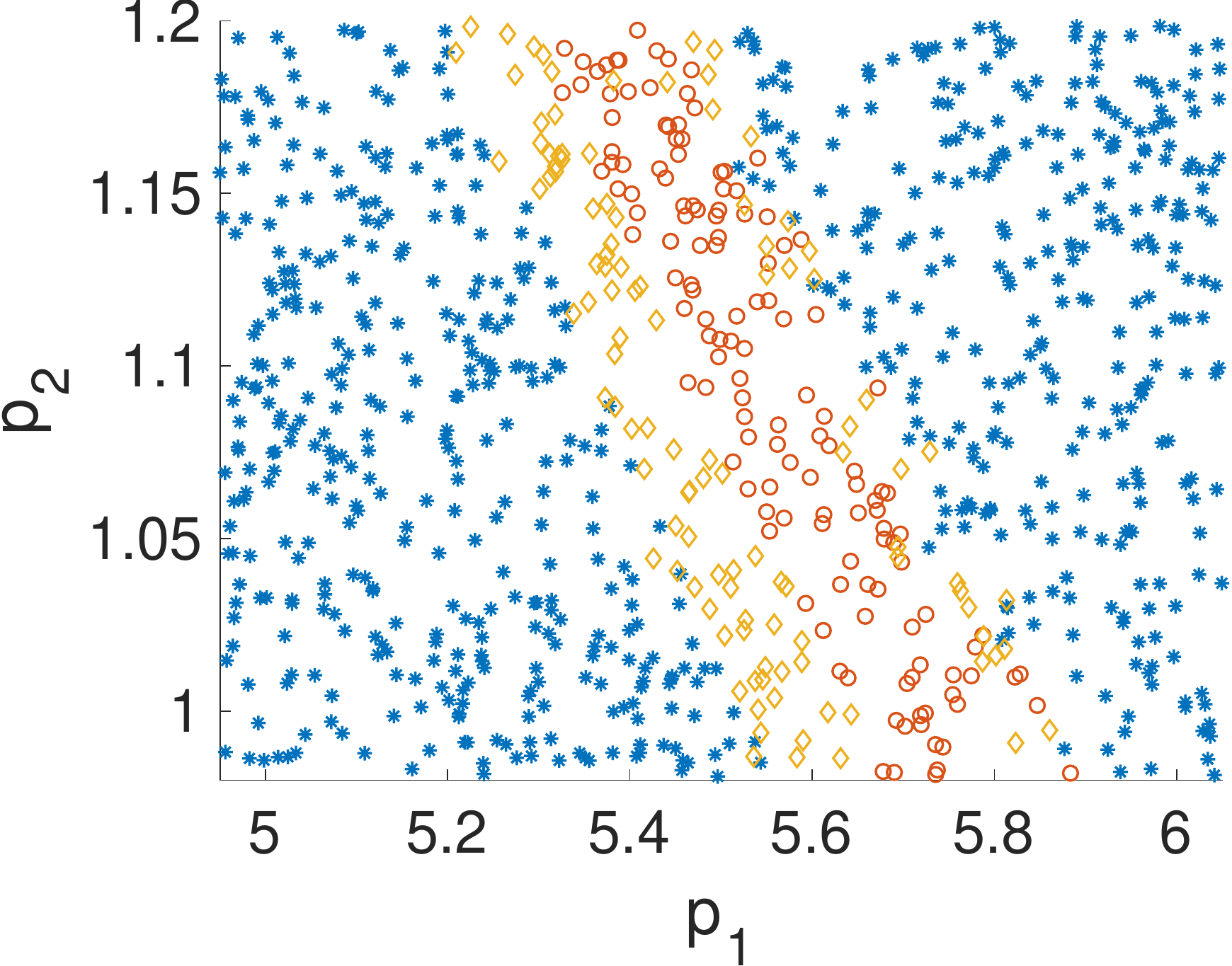}
\caption{Left: examples of stable (red) and unstable (black) steady state solutions. Right: samples of the buffer parameters $(\theta_{120},\theta_{121})$ using uniform independent sampling. A blue \textcolor{blue}{*} indicates the sample yielded a premature termination of the solver, a yellow \textcolor{yellow}{$\Diamond$} indicates the sample yielded an unstable steady state, a red \textcolor{red}{$\circ$} indicates the sample yielded a stable steady state.}
\label{steady_states}
\end{figure}

Applying the methodology described in Section~\ref{sec:param_dist_fit}, $S=919$ parameter samples are drawn and \eqref{caboodle} is solved for each sample with no stimulus applied. Of these 919 runs, 670 of them fail due to ode15s terminating prematurely because its time steps became too small. The remaining 249 yields 139 solutions for which the radius reaches a stable steady state (at least for the duration of the time integration) and 110 solutions for which the radius reaches a steady state but subsequently becomes transient. The left panel of Figure~\ref{steady_states} displays four representative solutions for the radius; the red curves remain in steady state for the duration of our time integration whereas the black curves revert into a transit regime after some time in or near steady state. The right panel of Figure~\ref{steady_states} displays the samples for two parameters (defined in the buffer equation \eqref{eqn:buff}, see the Supplementary Material for details) which are found to be highly influential in determining the behavior of the solution. A blue \textcolor{blue}{*} indicates a sample where the solver terminated prematurely, a yellow \textcolor{yellow}{$\Diamond$} indicates a sample where an unstable steady state was observed, a red \textcolor{red}{$\circ$} indicates a sample where the solution remained in steady state. A strong correlation determining the behavior of the solution is observed. The two parameters in question, denoted by ($\theta_{120},\theta_{121}$), occur in the buffer equation defined by
\begin{eqnarray}\label{eqn:buff}
[K^+]_e+B \quad\begin{array}{c}
f_f =\frac{\theta_{118}}{1+exp(\frac{-[K^+]_e+\theta_{120}}{\theta_{121}})}\\ 
 \rightleftharpoons \\
f_r=\theta_{118}
\end{array} 
[K^+]_b \nonumber \\
\frac{d[K^+]_b}{dt}=f_f[K^+]_e(\theta_{119}-[K^+]_b)-f_r[K^+]_b \nonumber \\
\end{eqnarray}
The parameters $\theta_{120}$ and $\theta_{121}$ have values 5.5 and 1.09, representing a shift and scaling in the forward rate constant, respectively. 

Observing this correlation, we use the accepted samples to fit $(\theta_{120},\theta_{121})$ with a bivariate Frank copula with beta marginals. The experiment is repeated by sampling the two correlated parameters from this bivariate distribution and all other parameter from their original uniform distributions. After four iterations refining the joint distribution of $(\theta_{120},\theta_{121})$, we were able to generate 902 out of 960 samples which yielded solutions with stable steady states (51 solutions had unstable steady states and 7 had premature solver terminations). This fitted distribution is used for all subsequent analysis. 

\begin{figure}[h]
\centering
\includegraphics[width=.3 \textwidth]{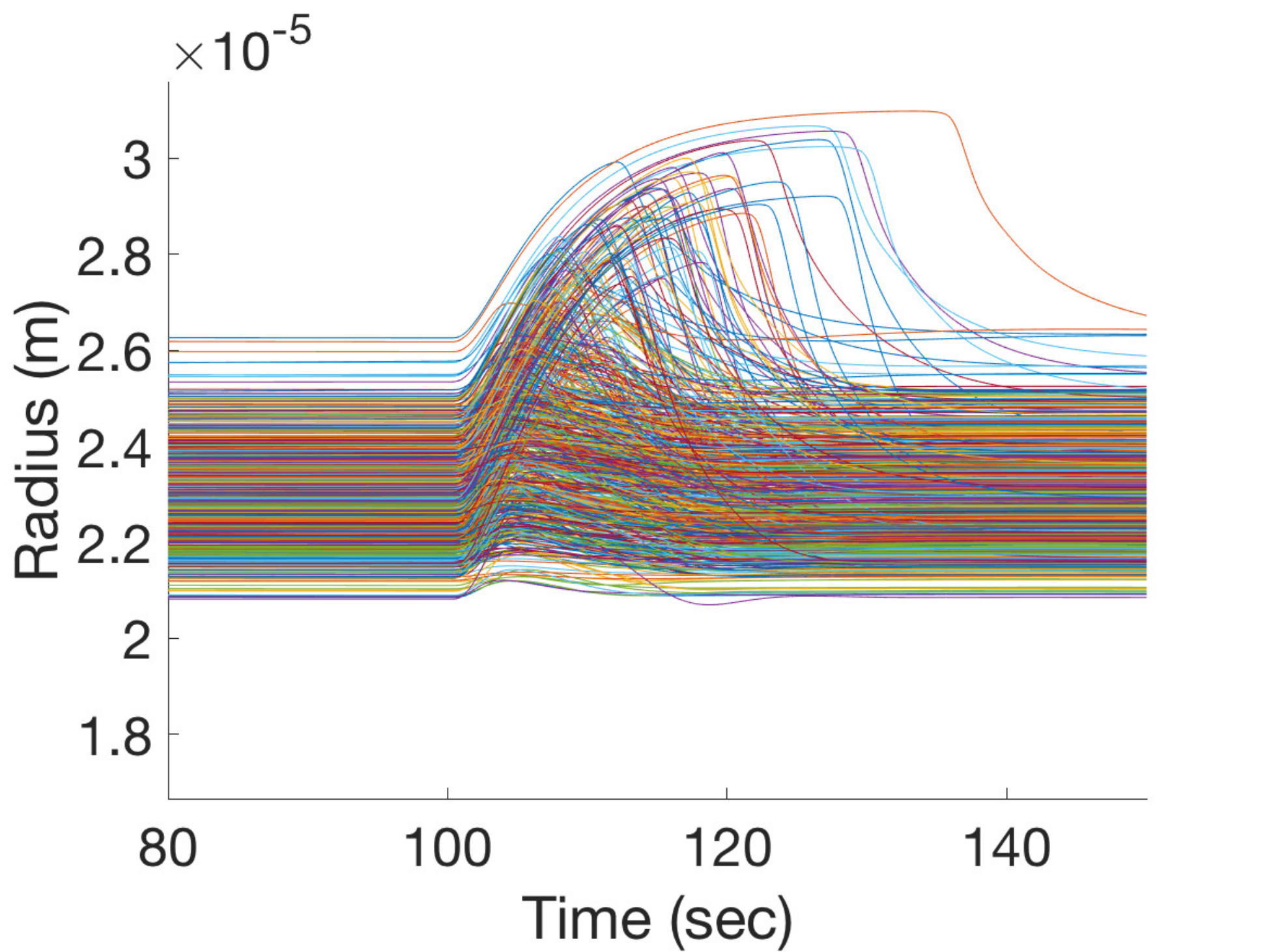}
\includegraphics[width=.3 \textwidth]{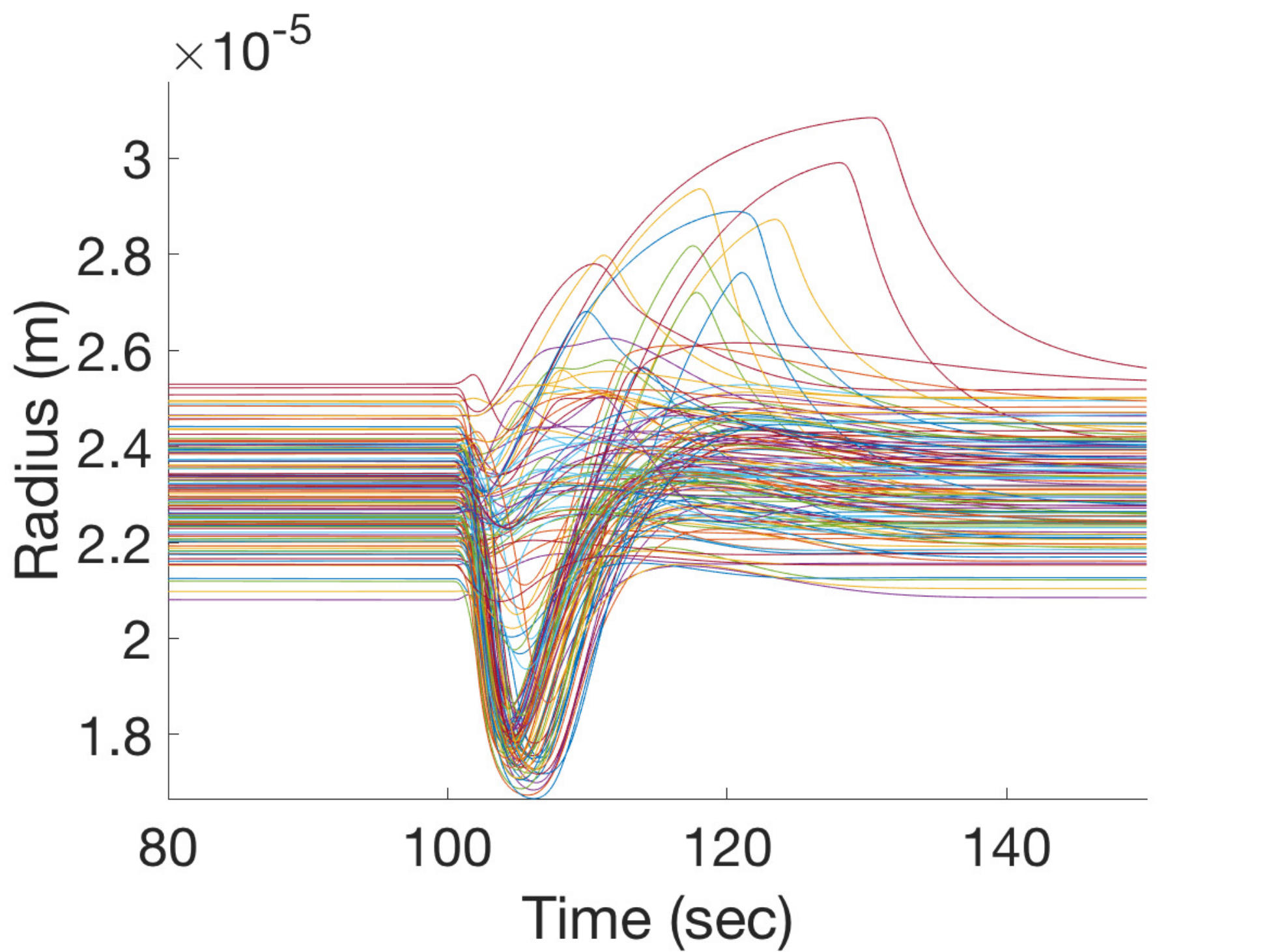}
\includegraphics[width=.3 \textwidth]{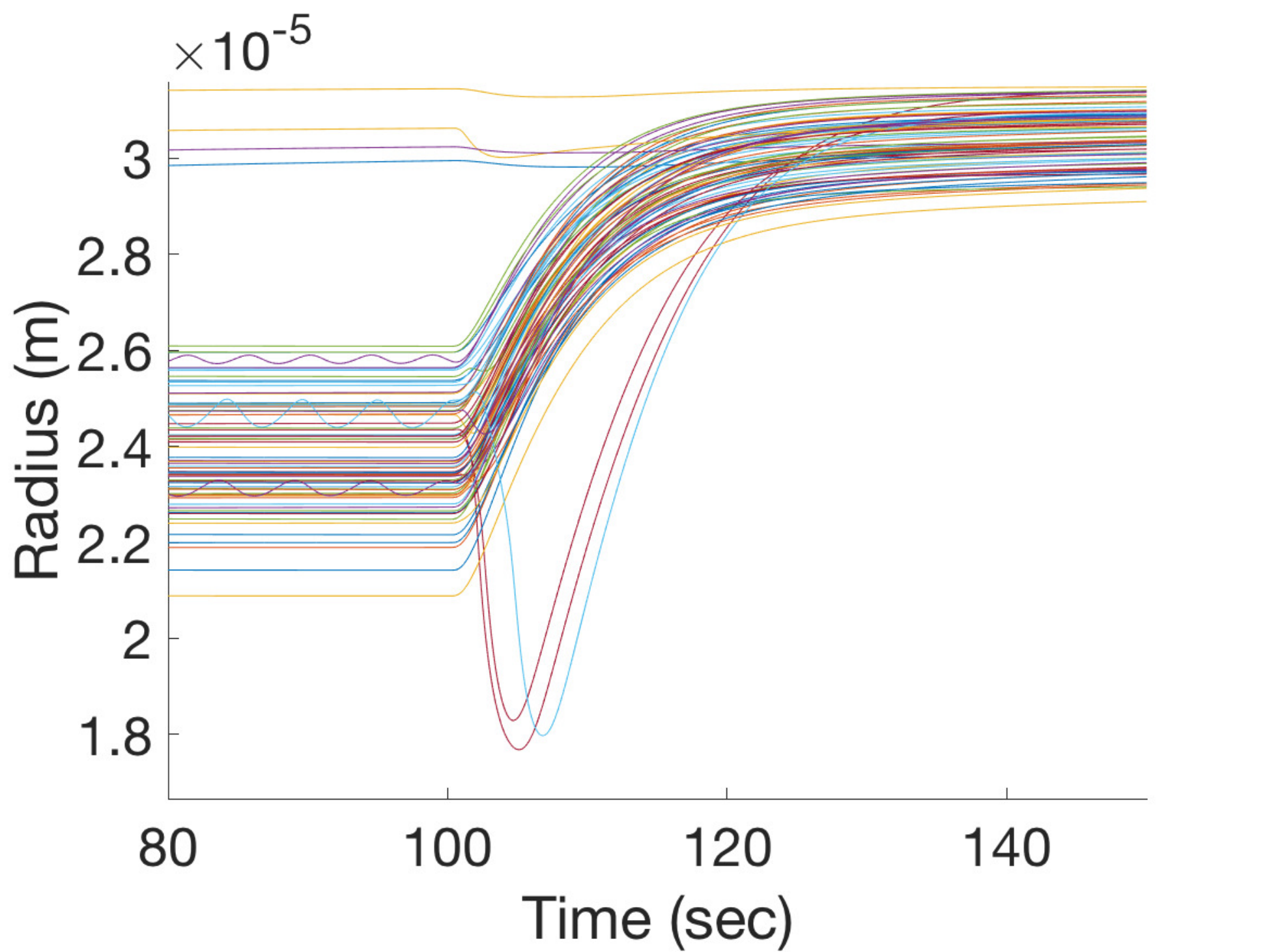}
\caption{Radii corresponding to samples (using the rectangular pulse stimulus). Left: curves an an increase in response to the stimulus; center: curves with a decrease in response to the stimulus; right: curves which settle in a different steady state.}
\label{solution_regimes}
\end{figure}

Samples are drawn and the model, with a stimulus applied (in two separate cases, the 10 second rectangular pulse and the 16 second experimental pulse), is run for each sample. This results in solutions exhibiting three different physiological regimes; they are displayed in Figure~\ref{solution_regimes} where the radius is plotted as a function of time. The leftmost panel corresponds to the typical case when the radius increases in response to the stimulus and then decreases when it is removed; the center panel corresponds to an atypical case where the radius has an initial decrease in response to the stimulus; the right panel corresponds to another atypical case where the radius reaches another steady state and does not decrease after the stimulus is removed.

This article focuses on the non-pathological case, corresponding to the left panel of Figure~\ref{solution_regimes}, so we remove samples where the radius does not increase in response to the stimulus and decrease when it is removed. However, we recognise that a decrease in radius upon stimulation does not necessarily mean an incorrect result. Indeed these cases are of particular importance (due to the possibility of the existence of cortical spreading depression \cite{Kenny2018b}) and a topic of future research. This processing yields 660 samples for analysis when the rectangular pulse stimulus is applied and 438 samples when the experimental pulse stimulus is applied. The results presented below use these samples.

Exploration of the 660 retained samples indicate that ``pathological'' cases have higher probability when the parameter which shifts the activation variable for the \pot flux through the soma KDR channel is reduced;
 however, this parameter does not characterize the solution regime by itself; it is likely that the solution regime is characterized by a combination of several parameters. Further sampling and exploration is required to better understand the structure in parameter space which determine the solution regime.

The 160 parameters are indexed (purely for coding reasons) and are not to be taken as a ranked order. The tables of ranked parameters, given in the subsections below to summarize the most influential parameters, provide the parameters in the first column, their location in the Supplementary Material in the second column, and their total Sobol' indices for the experimental and rectangular pulses in the third and fourth columns, respectively. Each table contains the five most influential parameters for a given QoI, as measured by the total Sobol' indices for the experimental pulse. There are slight differences in the ordering of the less important parameters for the rectangular pulse and experimental pulse cases; the tables below report the ranking from the experimental pulse case. Both rectangular and experimental pulse cases agree on the ranking of the most influential parameters.

\subsection{Average ECS Potassium}
\label{sec:qoi_K_ECS_Mean}

\begin{figure}[h]
\centering
\includegraphics[width=.46 \textwidth]{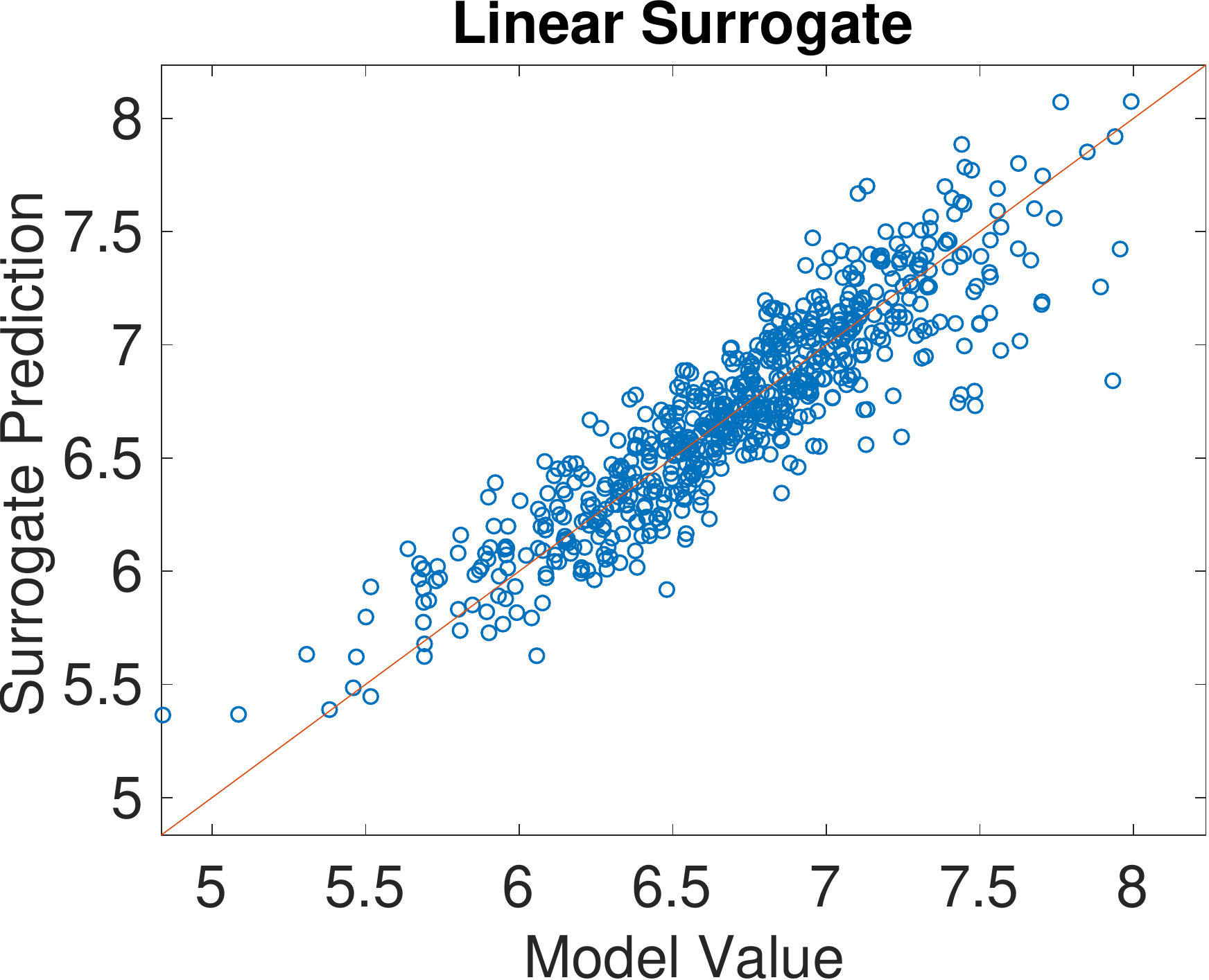}
\hspace{.1 cm}
\includegraphics[width=.475 \textwidth]{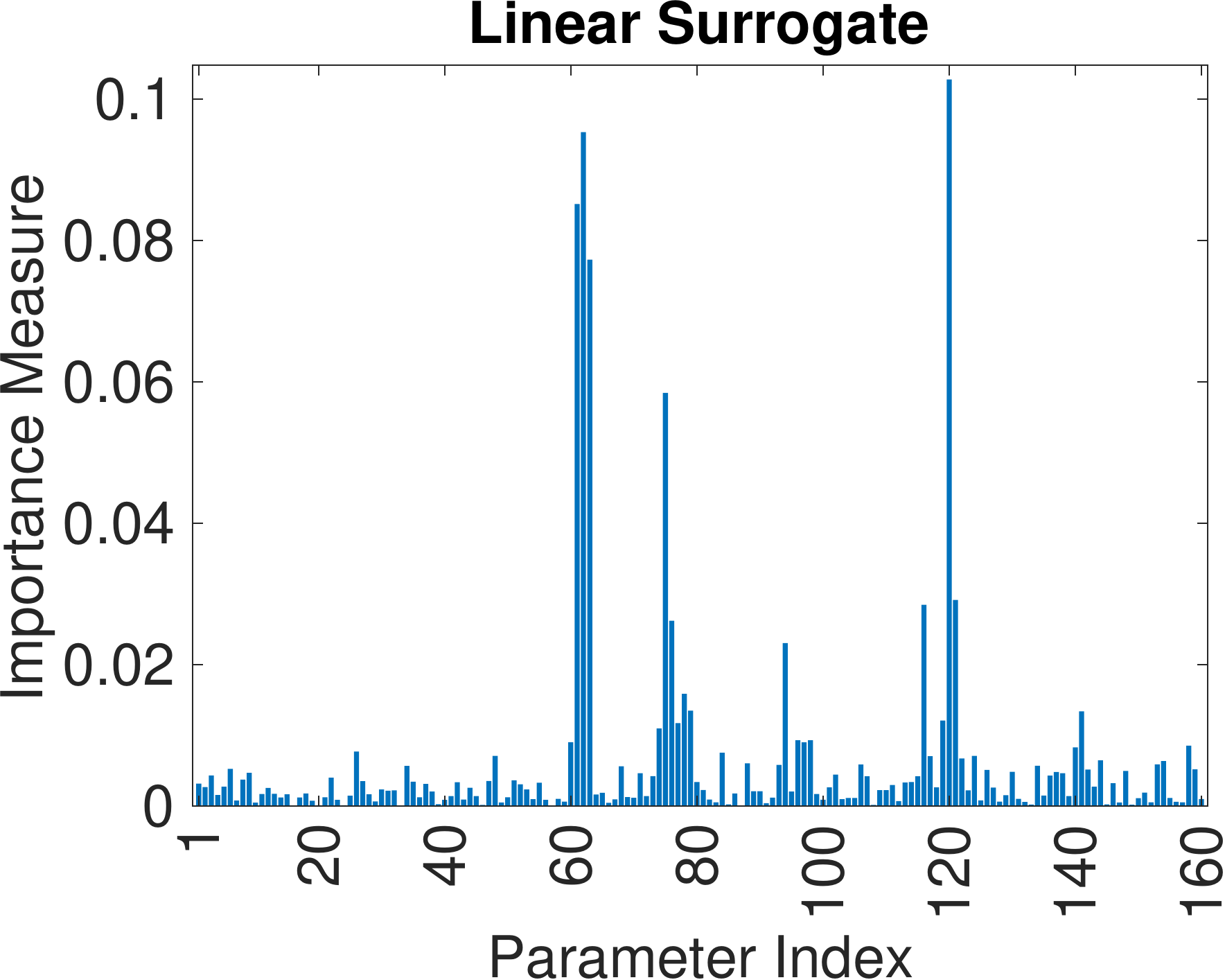} \\
\vspace{.2 cm}
\includegraphics[width=.46 \textwidth]{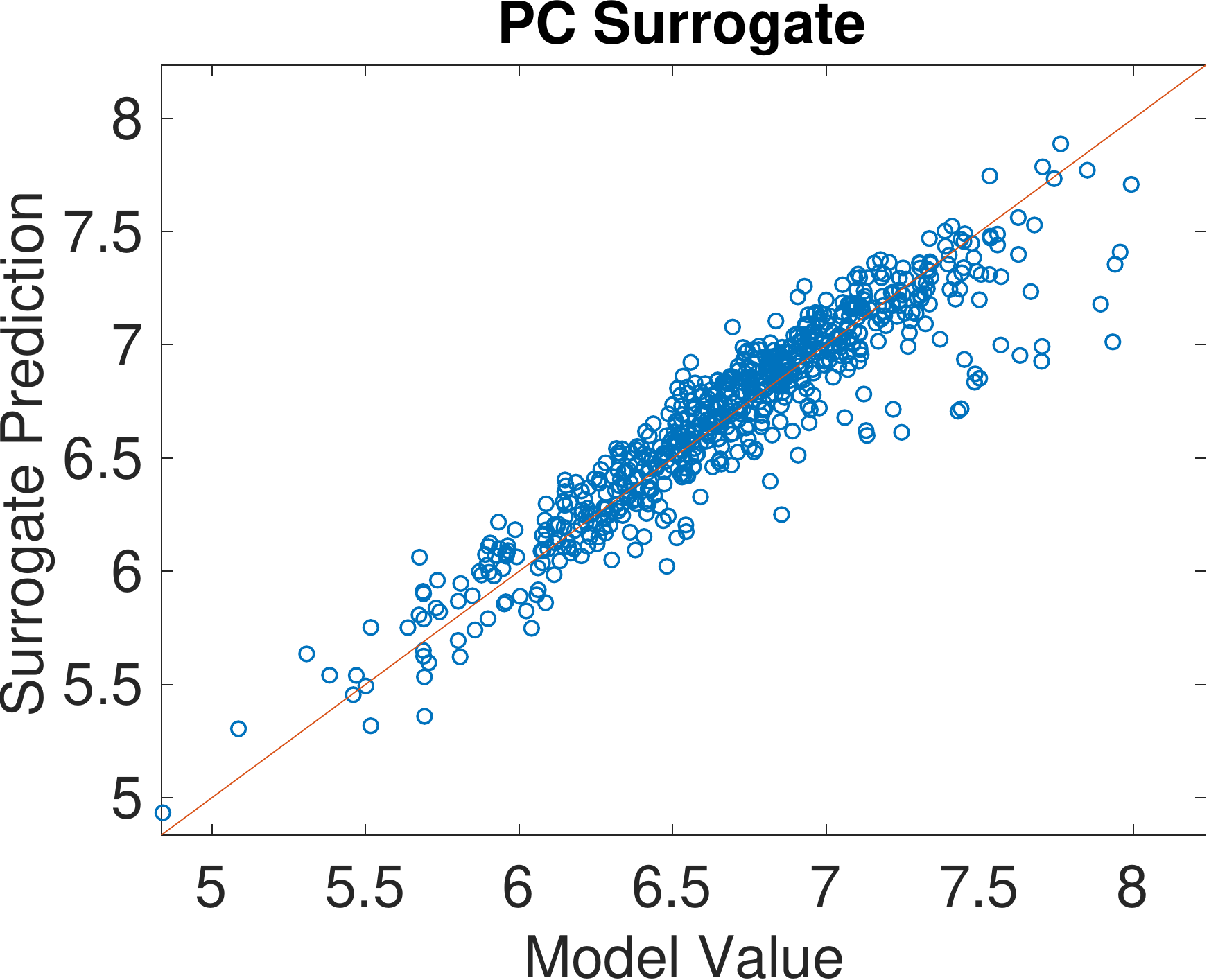}
\hspace{.1 cm}
\includegraphics[width=.475 \textwidth]{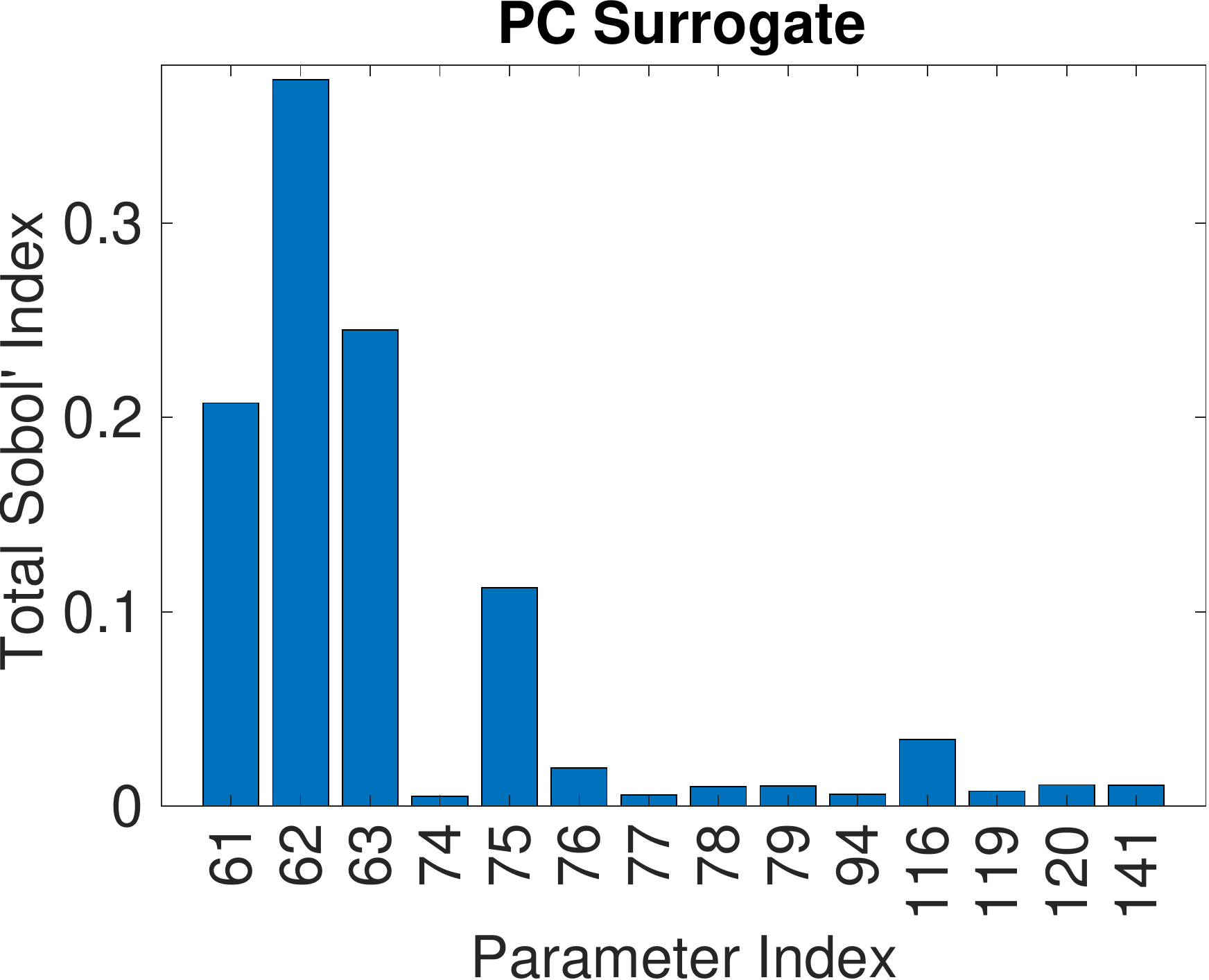}
\caption{Average ECS potassium QoI with a rectangular pulse stimulus. From left to right and top to bottom: linear surrogate predictions, linear surrogate importance measure, PC surrogate predictions, total Sobol' indices for PC surrogate.}
\label{fig:K_ECS_Mean_rect}
\end{figure}

\begin{figure}[h]
\centering
\includegraphics[width=.46 \textwidth]{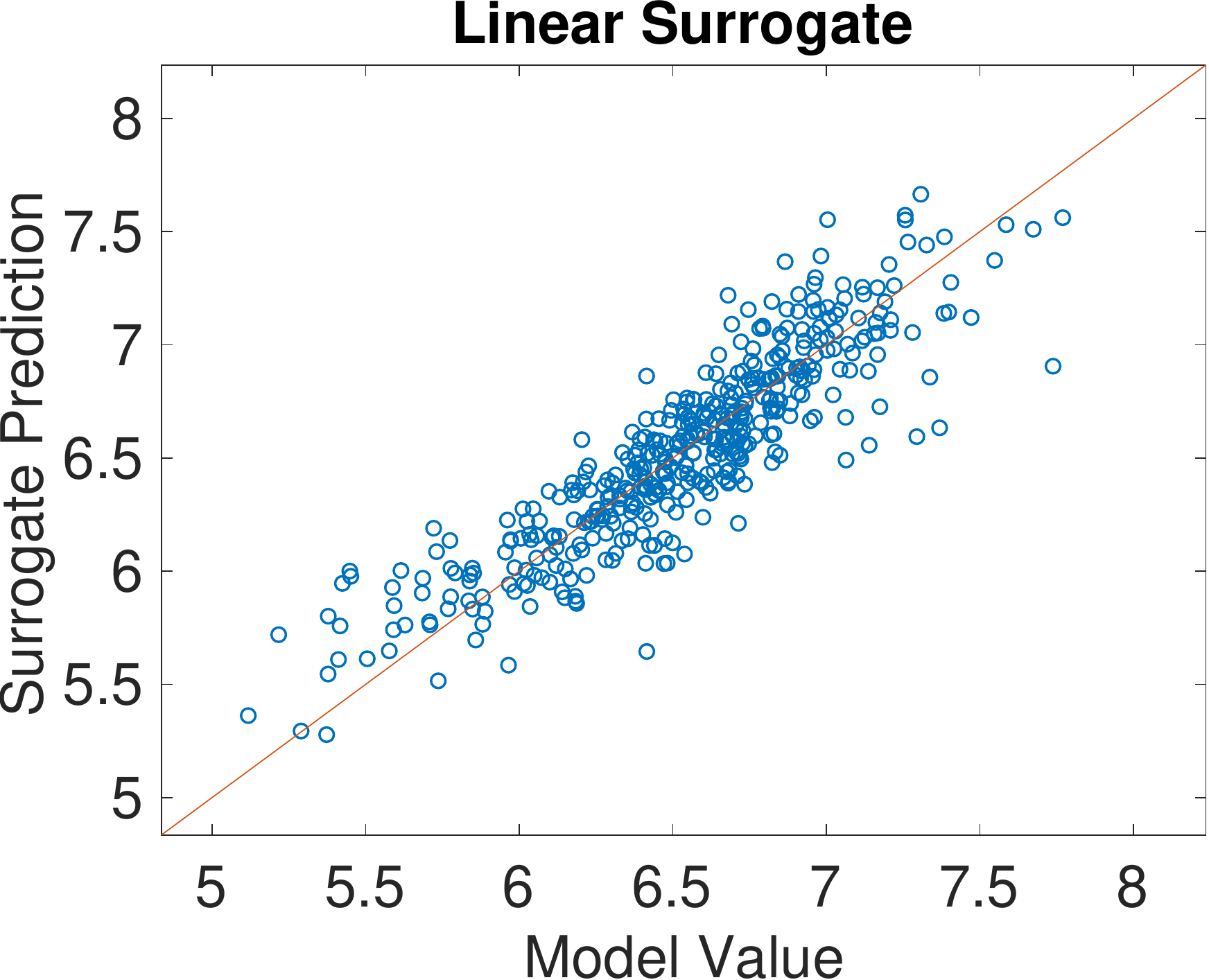}
\hspace{.1 cm}
\includegraphics[width=.475 \textwidth]{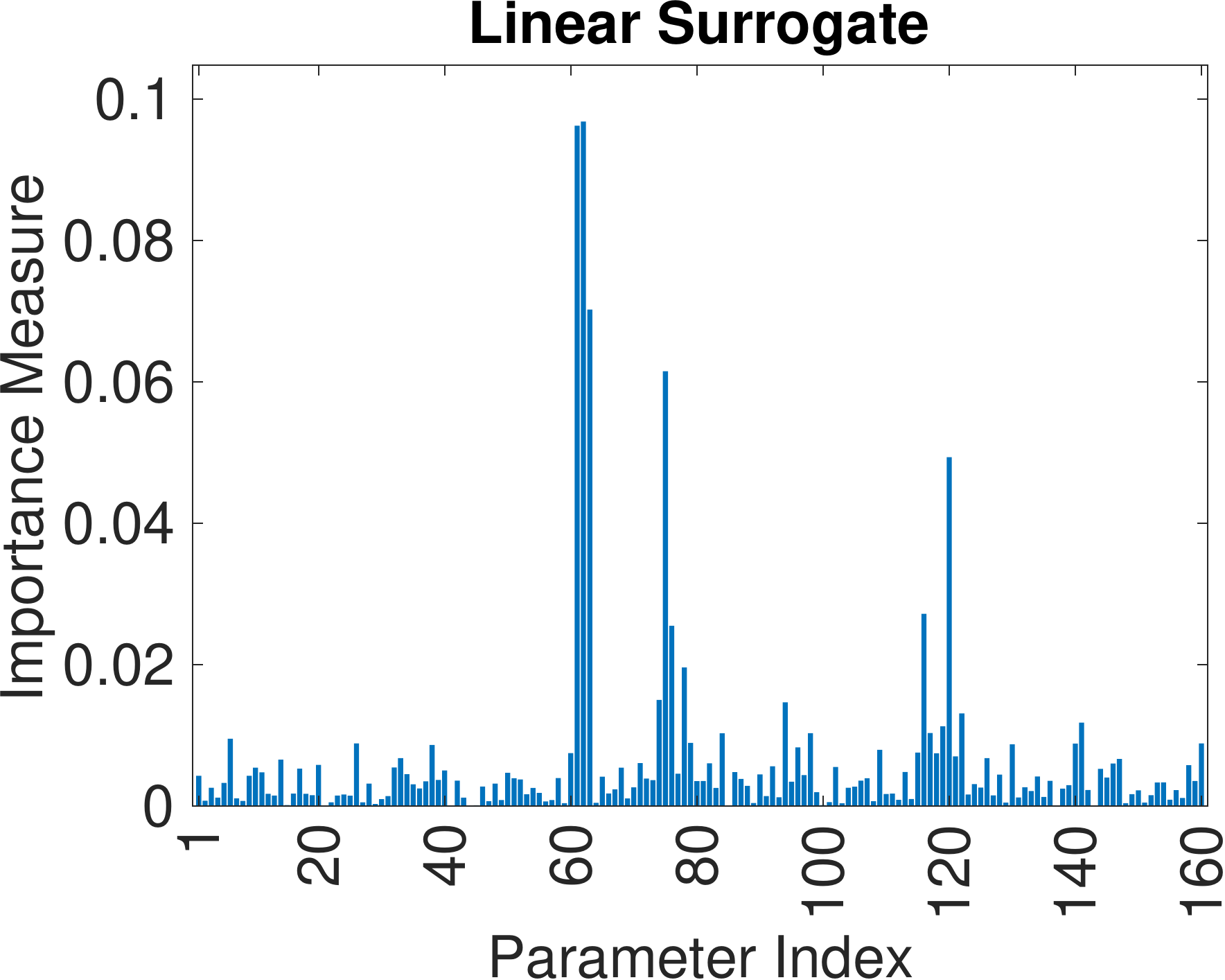} \\
\vspace{.2 cm}
\includegraphics[width=.46 \textwidth]{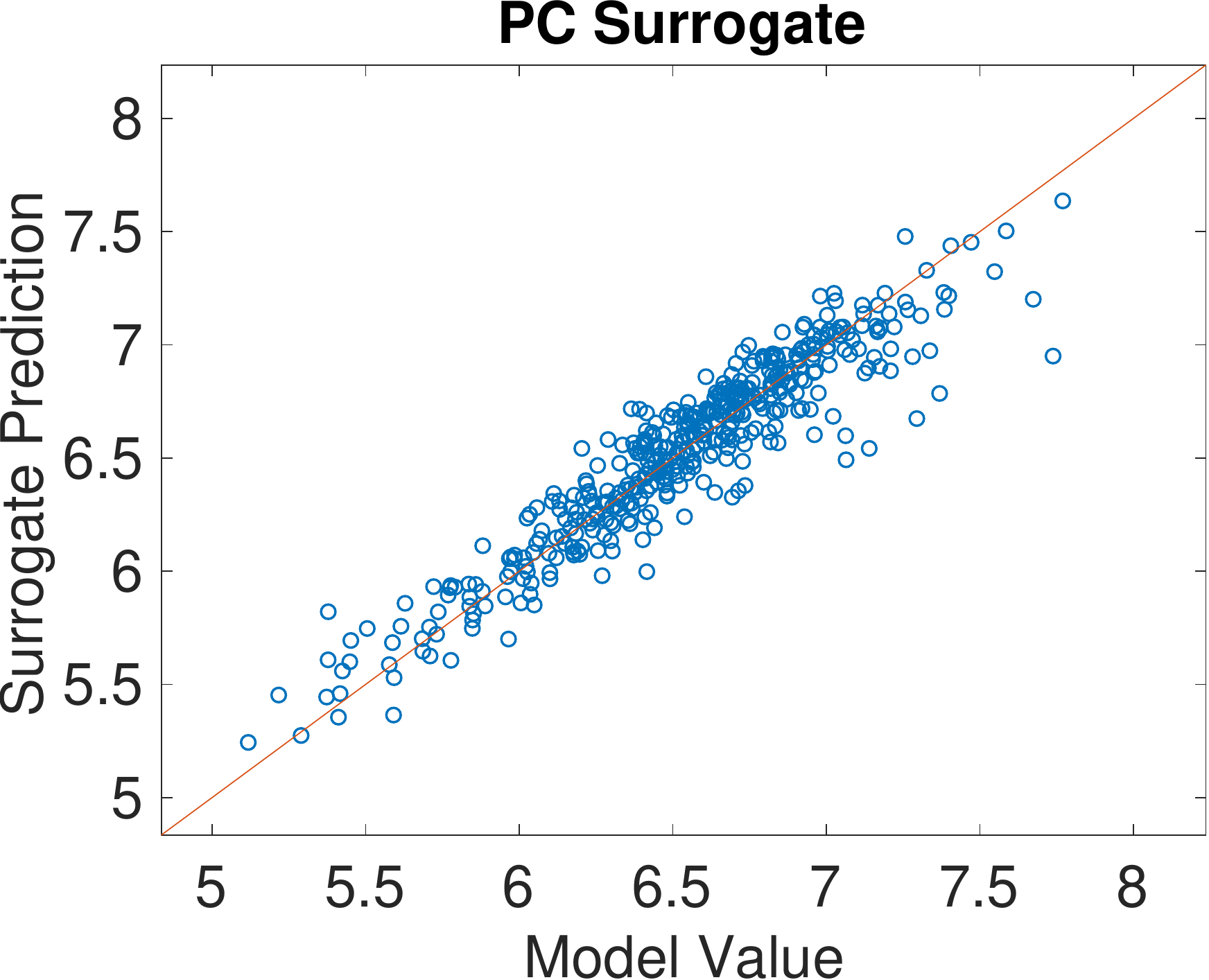}
\hspace{.1 cm}
\includegraphics[width=.475 \textwidth]{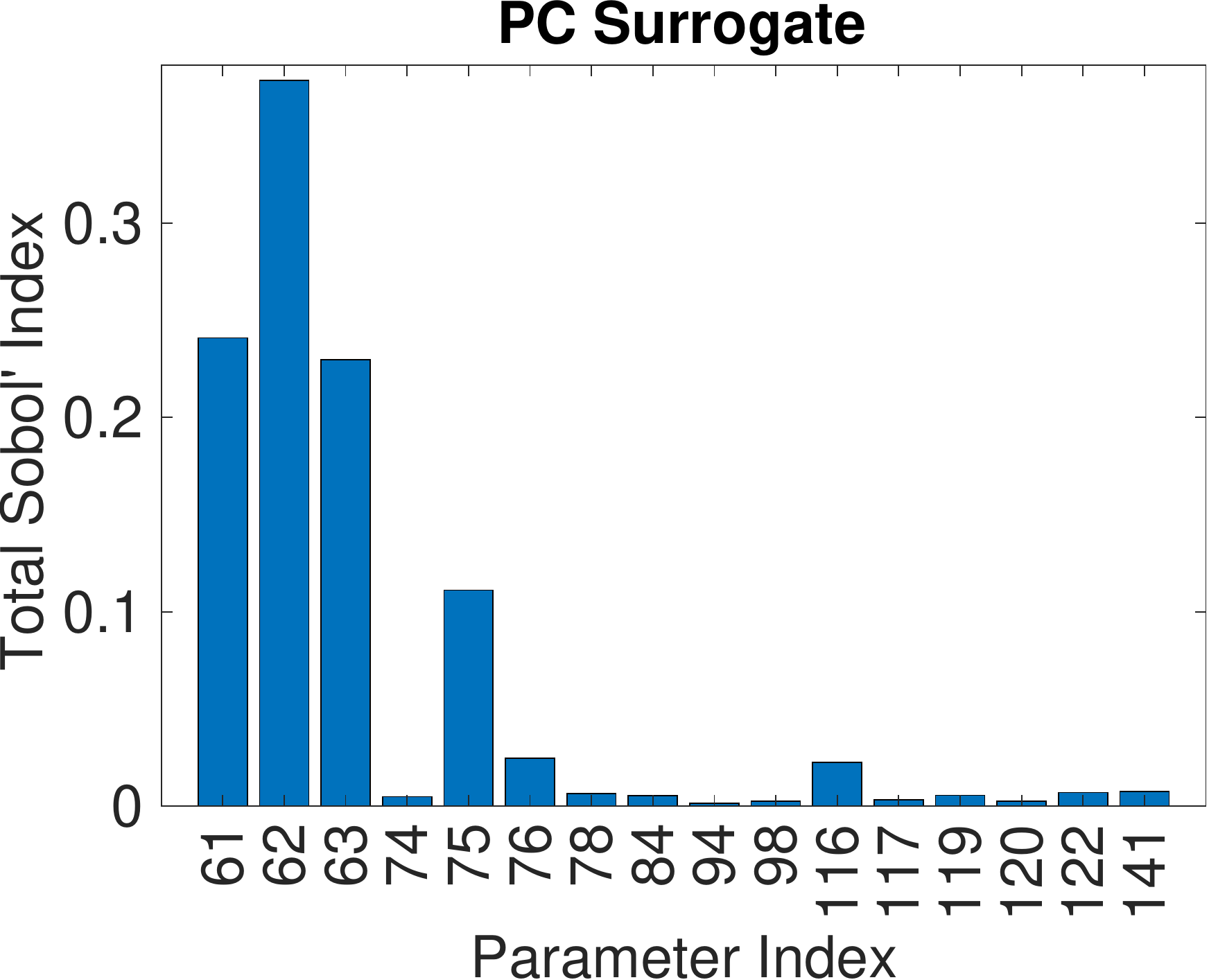}
\caption{Average ECS potassium QoI with experimental pulse stimulus. From left to right and top to bottom: linear surrogate predictions, linear surrogate importance measure, PC surrogate predictions, total Sobol' indices for PC surrogate.}
\label{fig:K_ECS_Mean_exp}
\end{figure}

Figure~\ref{fig:K_ECS_Mean_rect} and ~\ref{fig:K_ECS_Mean_exp} display results for the average of the ECS potassium, defined in equation (\ref{K_ECS_Mean}), for the rectangular pulse stimulus and experimental pulse stimulus, respectively. In the top left panels, predictions of the linear surrogate are plotted against the QoI values. The sensitivities $L_j$, $j=1,2,\dots,160$, are displayed in the top right panels. Predictions of the PC surrogate are plotted against the QoI values in the bottom left panels. The total Sobol' indices of the PC surrogate are given in the bottom right panels. Table~\ref{tab:K_ECS_Mean} reports the five most important parameters and their respective total Sobol' indices.


\begin{table}[h]
\centering
\ra{1.3}
\begin{tabular}{cccc}
\toprule
Parameter & Identification in Supplementary Material & Total Sobol' Index (exp.) & Total Sobol' Index (rect.)\\
\midrule
$\theta_{62}$ & Nominal value 0.143 in equation (61) &  0.3732 & 0.3738\\
$\theta_{61}$ & $g_{K,leak_d}$ in equation (30) & 0.2409 & 0.2074\\
$\theta_{63}$ &  Nominal value 5.67 in equation (61) &  0.2297 & 0.2449 \\
$\theta_{75}$ & Nominal value 34.9 in equation (54) & 0.1112 & 0.1124\\
$\theta_{76}$ & Nominal value 0.2 in equation (54) & 0.0247 & 0.0198\\
 \arrayrulecolor{black}\bottomrule
\end{tabular}
\caption{Five most influential parameters for the average ECS potassium QoI when the experimental pulse stimulus is applied. The leftmost column is the parameter, the left-center column identifies the parameter in the Supplementary Material, the right-center column is the total Sobol' index computed for the parameter using the experimental pulse stimulus, and the right column is the total Sobol' index computed for the parameter using the rectangular pulse stimulus.}
\label{tab:K_ECS_Mean}
\end{table}
The first and third parameters in Table \ref{tab:K_ECS_Mean} are scaling and shift parameters ($\theta_{62}$ and $\theta_{63}$) for the activation gating variable, $m_4$ in the dendrite NaP channel respectively, whose ODE is defined as 
\begin{eqnarray}\label{eqn:m4}
\frac{dm_4}{dt}=m_{4 \alpha}(1-m_4)-m_{4 \beta}m_4, \nonumber \\
m_{4 \alpha}=  \frac{1}{6(1 + exp(-(\theta_{62}  v_d + \theta_{63})))},\nonumber \\
m_{4 \beta}= \frac{exp(-(\theta_{62} v_d + \theta_{63}))}{6(1 + exp(-(\theta_{62} v_d + \theta_{63})))}.
\end{eqnarray}
where $v_d$ is the dendrite membrane potential. \\
The nominal values of $\theta_{62}$ and $\theta_{63}$ are 0.143 and 5.67, respectively. These effectively define the characteristic time scale and forcing function in the rate equation for the open  probability of the persistent sodium channel. The second most important parameter determines the strength of the  conductance in the \pot leak ion channel. The fourth and fifth parameters, $\theta_{75}$ and $\theta_{76}$, are the shift and scale of the neuron membrane potential in the ODE for the activation variable for the  K flux through dendritic KDR channel, defined as 
\begin{eqnarray}\label{eqn:m6}
m_{6  \alpha}     = \theta_{74} \left(\frac{v_d + \theta_{75}}{1 - exp(-(\theta_{76}  v_d + \theta_{76} \theta_{75}))} \right), 
\end{eqnarray}
where the nominal values for $\theta_{74}, \theta_{75}$, and $\theta_{76}$ are 0.016, 34.9, and 0.2, respectively. 

The results for this specific QoI are similar for both the rectangular pulse and experimental pulse stimulus. In both cases, the QoI is approximated with reasonable accuracy by a linear surrogate and with higher accuracy by the PC surrogate. The most important parameters are shared in both cases. Notice that parameter $\theta_{120}$, defined in \eqref{eqn:buff}, appears to be important in the linear surrogate but unimportant in the PC surrogate. This is because it is strongly correlated with parameter $\theta_{121}$ (also defined in defined in \eqref{eqn:buff}), see Figure~\ref{steady_states}, and as a result the coefficient in the linear surrogate may be very large because its effect is offset by the effect of $\theta_{121}$. To decorrelate inputs, the PC surrogate is build with only $\theta_{120}$ instead of both $\theta_{120}$ and $\theta_{121}$. It subsequently has minimal importance.

We also analysed another QoI for the ECS potassium, namely, its maximum over the interval of stimulation. The results are not reported because of their similarity to the average ECS potassium QoI results given above.

The remaining two QoIs also present very similar results for both the rectangular pulse and experimental pulse stimulus. In the interest of conciseness, we only present figures corresponding to the experimental pulse stimulus for these two QoIs; Tables~\ref{tab:qoi_vol_flow} and ~\ref{tab:qoi_AM_AMp_Min} give results for both stimuli. 

\subsection{Average Volumetric Flow Rate}
\begin{figure}[h!]
\centering

\includegraphics[width=.46 \textwidth]{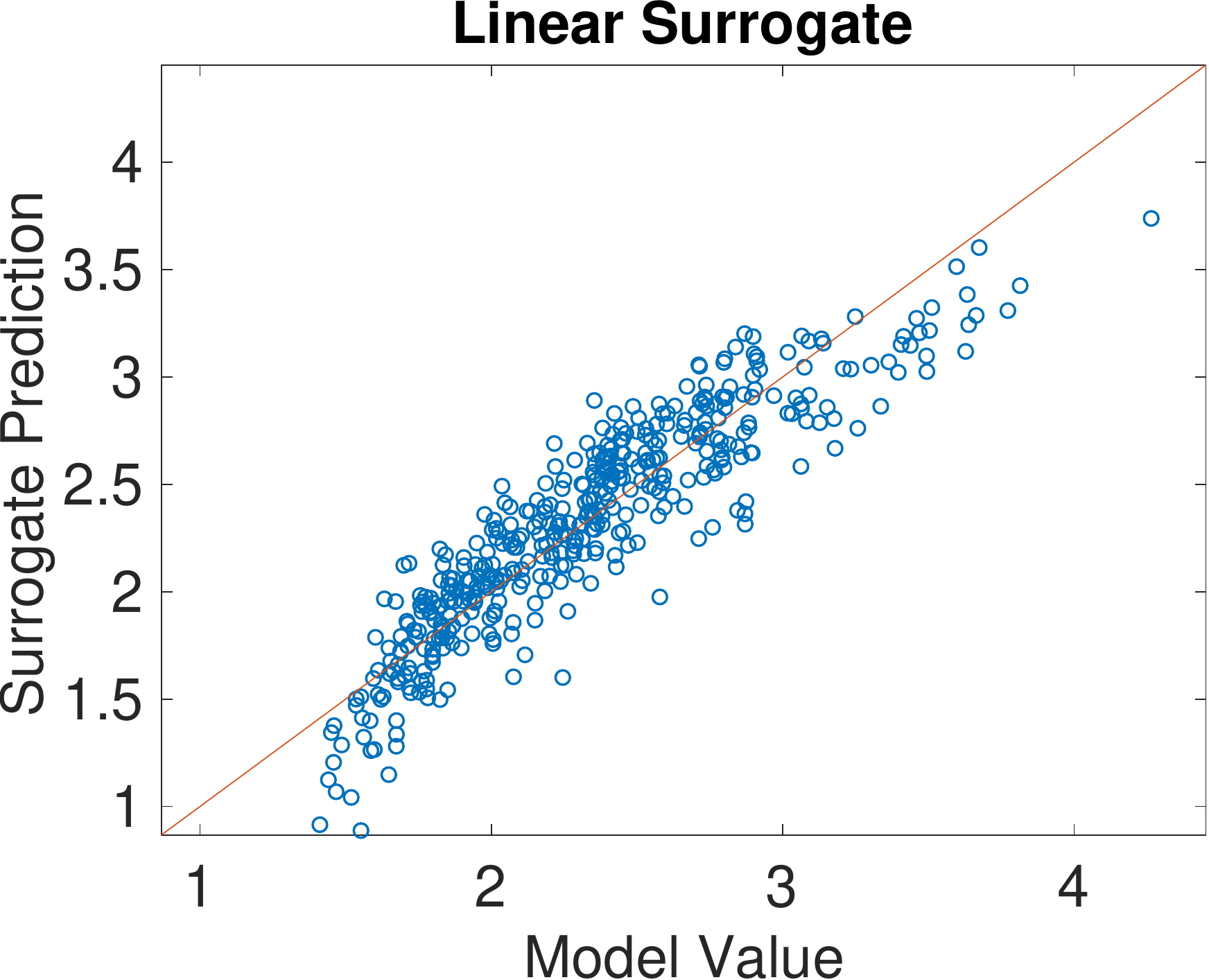}
\hspace{.1 cm}
\includegraphics[width=.475 \textwidth]{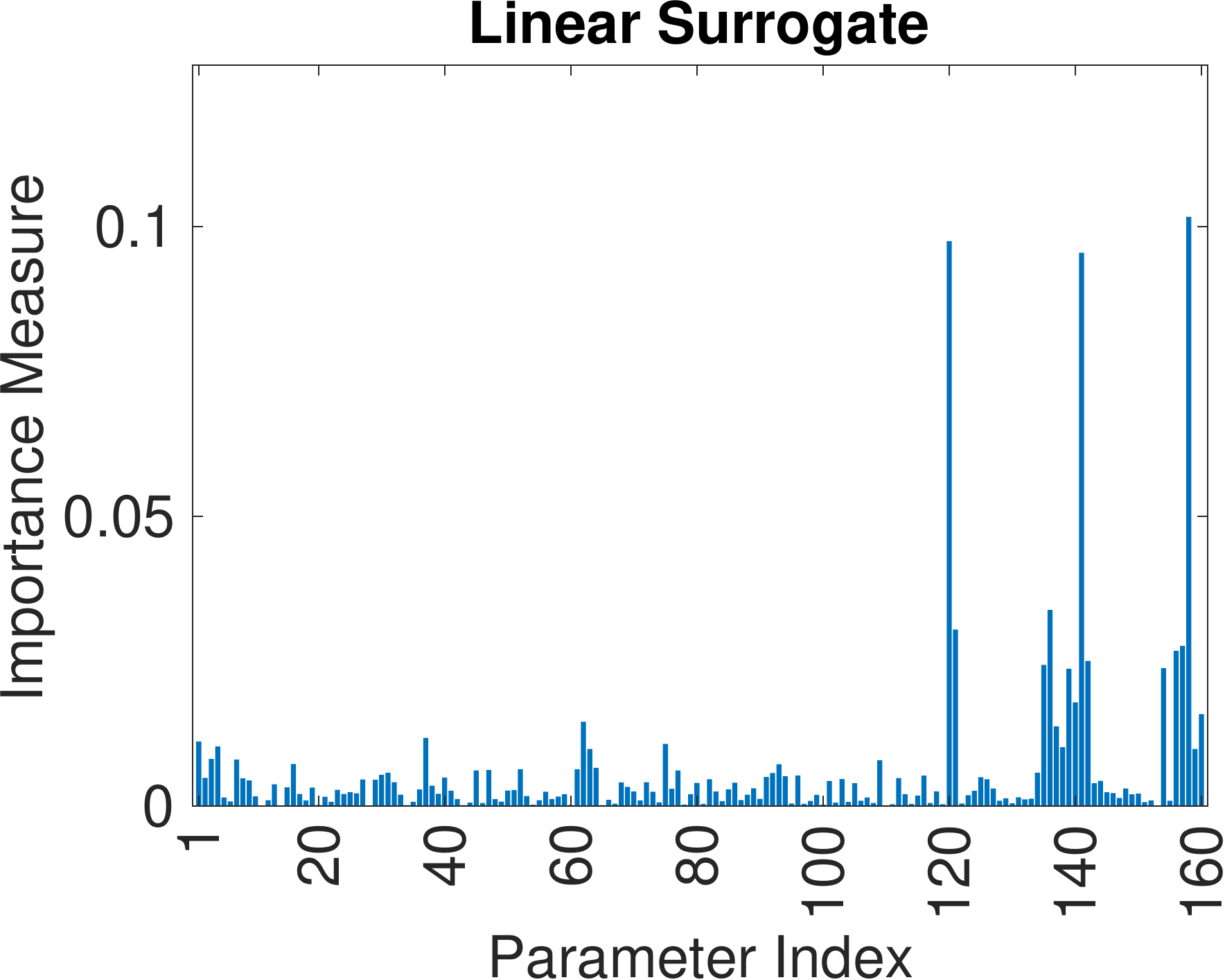} \\
\vspace{.2 cm}
\includegraphics[width=.46 \textwidth]{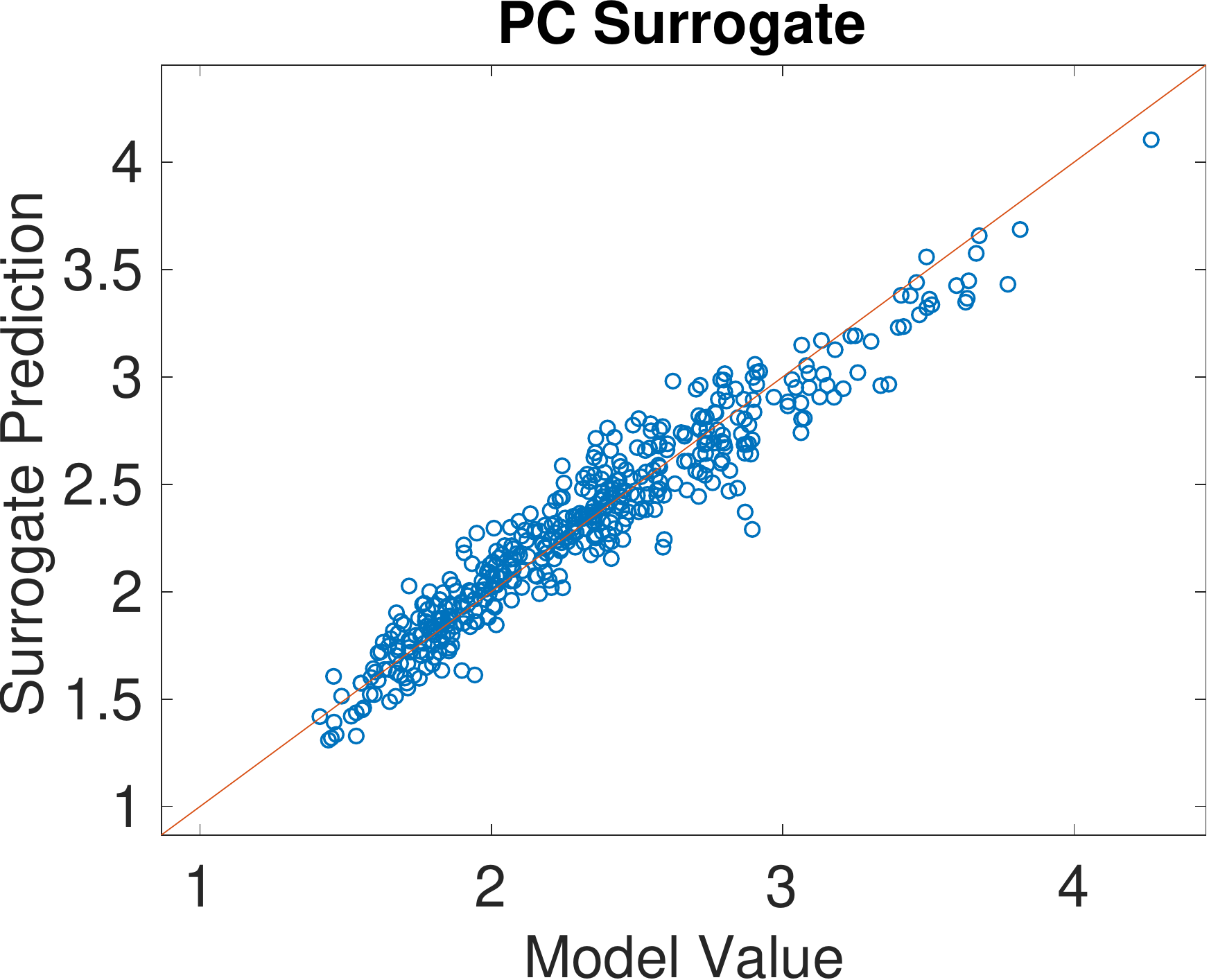}
\hspace{.1 cm}
\includegraphics[width=.475 \textwidth]{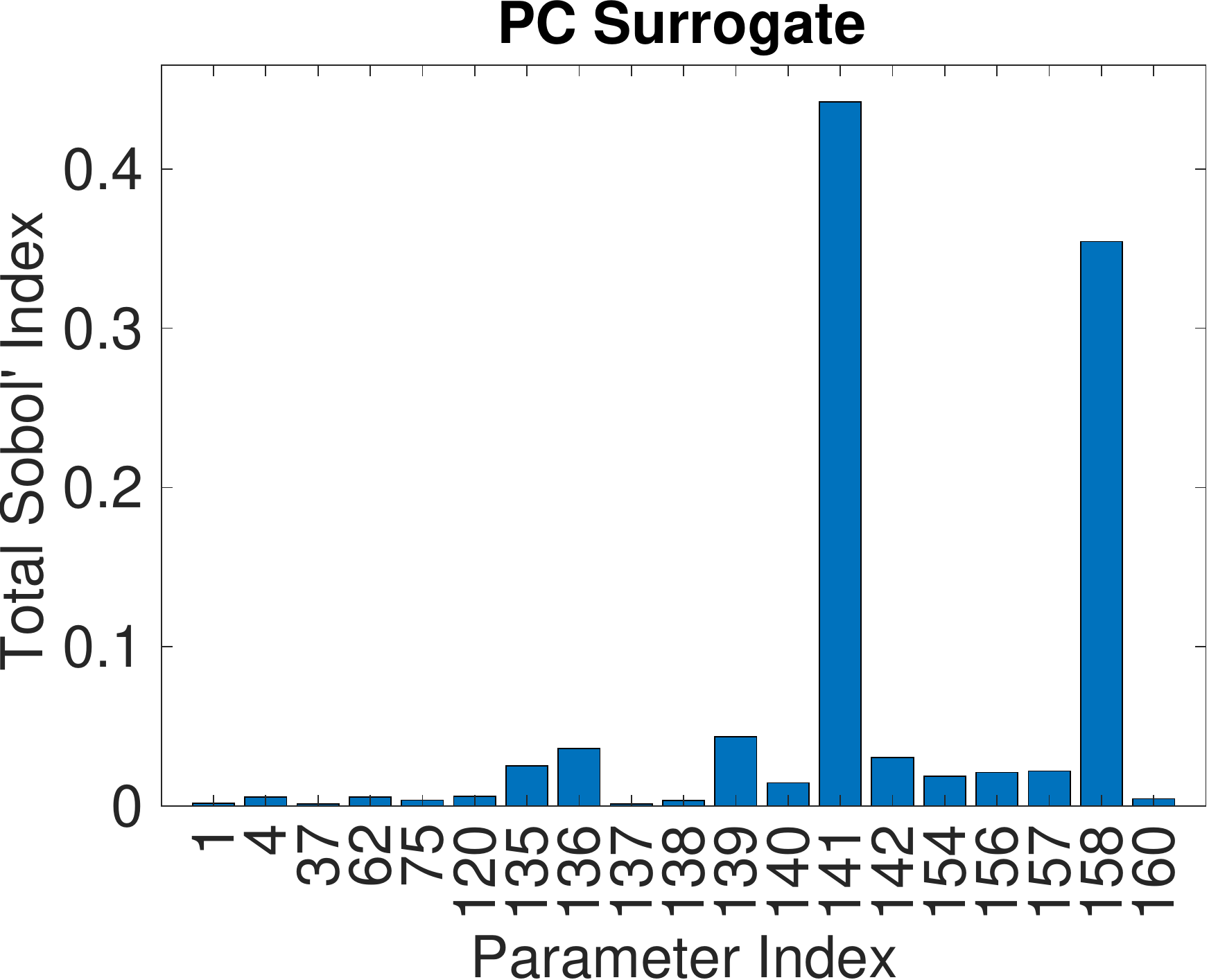}
\caption{Average volumetric flow rate QoI with experimental pulse stimulus. From left to right and top to bottom: linear surrogate predictions, linear surrogate importance measure, PC surrogate predictions, total Sobol' indices for PC surrogate.}
\label{fig:qoi_vol_flow_exp}
\end{figure}

Figure~\ref{fig:qoi_vol_flow_exp} displays results for the average volumetric flow rate in the cerebral tissue defined by (\ref{vol_flow}) in the same manner as Figures~\ref{fig:K_ECS_Mean_rect} and \ref{fig:K_ECS_Mean_exp}.   Table~\ref{tab:qoi_vol_flow} reports the five most important parameters and their total Sobol' indices. Unsurprisingly, the parameter list contains values found in the SMC/EC compartment of the full model. However, the topmost parameter, $\theta_{141}$, is associated with the conductance of the inwardly rectifying SMC KIR channel, $g_{KIR}$, defined as a function of both membrane potential $v_{SMC}$ and the \pot concentration in the perivascular space $[K^+]_{PVS}$, given by 
\begin{eqnarray}
g_{KIR}=exp\left( \theta_{142} v_{SMC}+\theta_{140}[K^+]_{PVS}-\theta_{141}\right)  \label{eqn:gkir}
\end{eqnarray}
as shown in \cite{Dormanns2015} fitting to the data of \cite{Filosa2006}. $\theta_{141}$ shifts the conductance to the right for constant $[K^+]_{PVS}$  concentration in the perivascular space. The second parameter in Table \ref{tab:qoi_vol_flow}, $\theta_{158}$ (found in the wallmechanics section of the model) determines the strength of influence of cytosolic $[Ca^{2+}]$ in determining the reaction rate of phosphorylation of myosin \cite{Hai1988}. Although not especially important, the third listed parameter $\theta_{139}$ shifts the Nernst potential for the KIR channel to the right in the equation
\begin{eqnarray}
v_{KIR}=\theta_{138} [K^+]_{PVS}-\theta_{139}. \label{vkir}
\end{eqnarray}

We again observe reasonably accurate fits by a linear surrogate and improved accuracy by a PC surrogate. The surrogate predictions and important parameters for the rectangular pulse and experimental pulse closely agree.  As in Subsection~\ref{sec:qoi_K_ECS_Mean}, parameter $\theta_{120}$ appears important in the linear surrogate but unimportant in the PC surrogate.

\begin{table}[h]
\centering
\ra{1.3}
\begin{tabular}{cccc}
\toprule
Parameter & Identification in Supplementary Material & Total Sobol' Index (exp.) & Total Sobol' Index (rect.) \\
\midrule
$\theta_{141}$ &  $z_4$ in equation (149)  & 0.4420 & 0.4561\\
$\theta_{158}$ & $n_{cross}$ in equation (214)   & 0.3544 & 0.3311\\ 
 $\theta_{139}$ & $z_2$ in equation (148)  & 0.0436 & 0.0529\\
$\theta_{136}$ & $G_{K_i}$ in equation (149)   & 0.0362 & 0.0305\\
$\theta_{142}$ & $z_5$ in equation (149)   &  0.0305 & 0.0351\\
   \arrayrulecolor{black}\bottomrule
\end{tabular}
\caption{Five most influential parameters for the average volumetric flow rate QoI when the experimental pulse stimulus is applied. The leftmost column is the parameter, the left-center column identifies the parameter in the Supplementary Material, the right-center column is the total Sobol' index computed for the parameter using the experimental pulse stimulus, and the right column is the total Sobol' index computed for the parameter using the rectangular pulse stimulus.}
\label{tab:qoi_vol_flow}
\end{table}

\subsection{$[AM+AM_p]_{min}$}

\begin{figure}[h]
\centering

\includegraphics[width=.46 \textwidth]{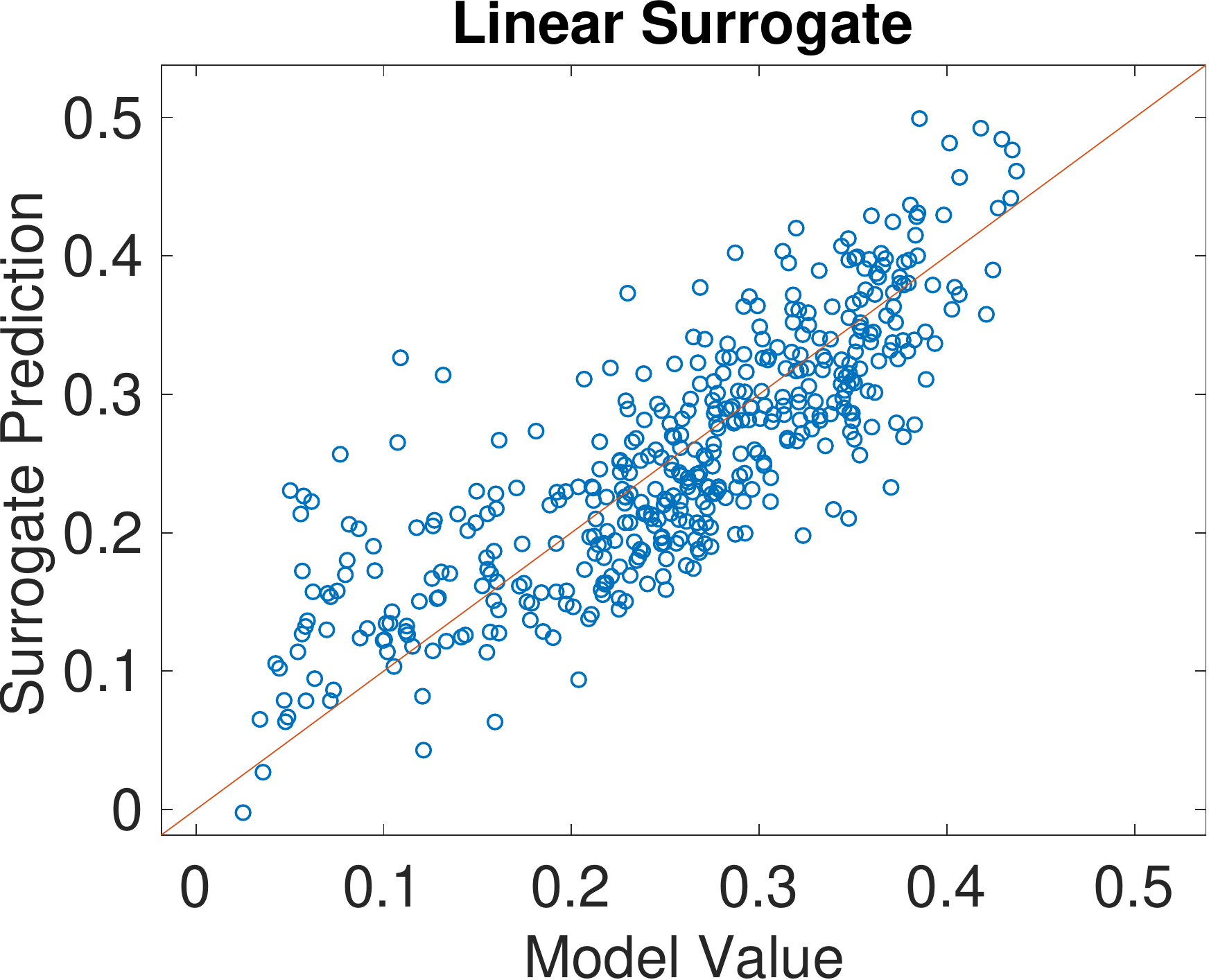}
\hspace{.1 cm}
\includegraphics[width=.475 \textwidth]{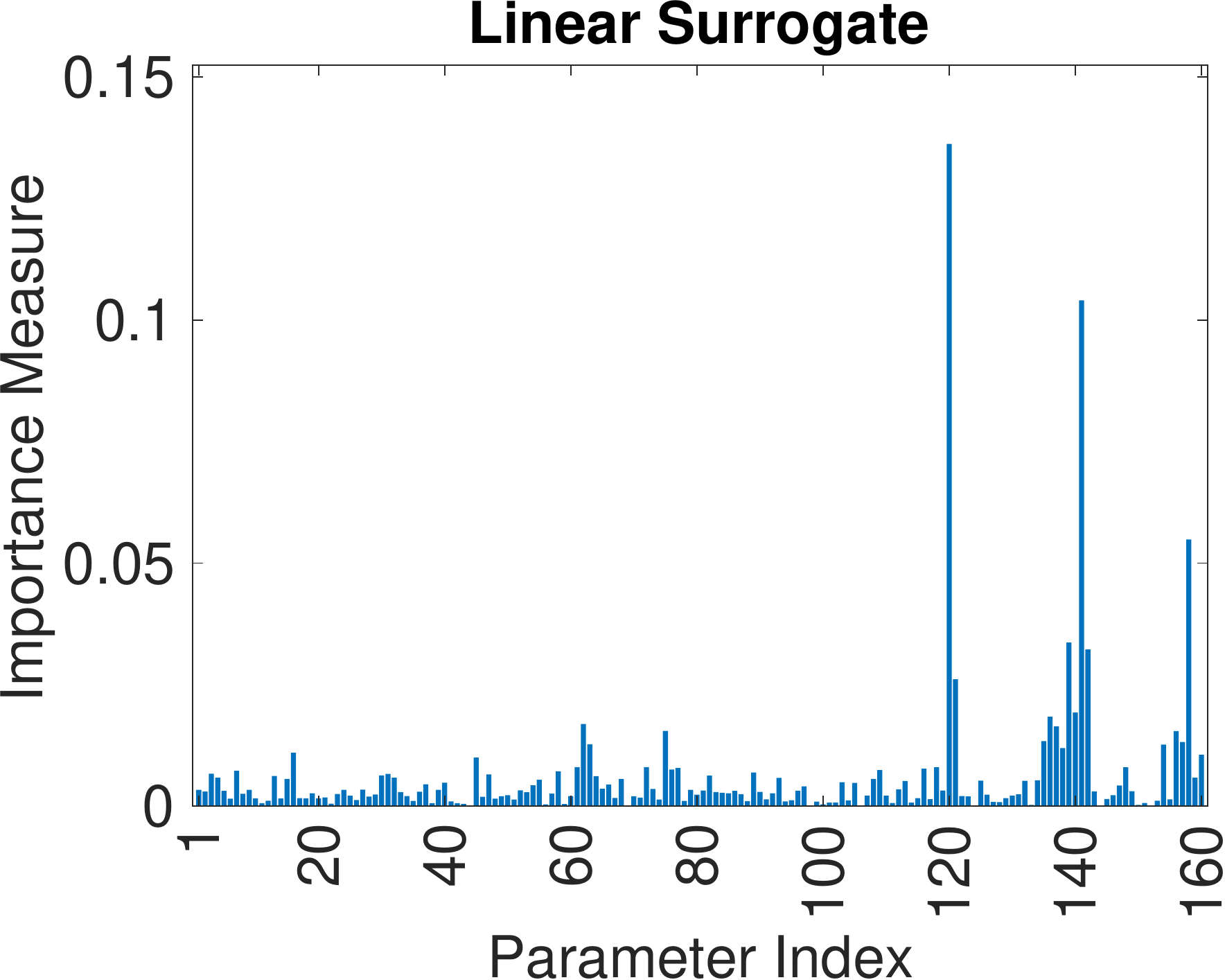} \\
\vspace{.2 cm}
\includegraphics[width=.46 \textwidth]{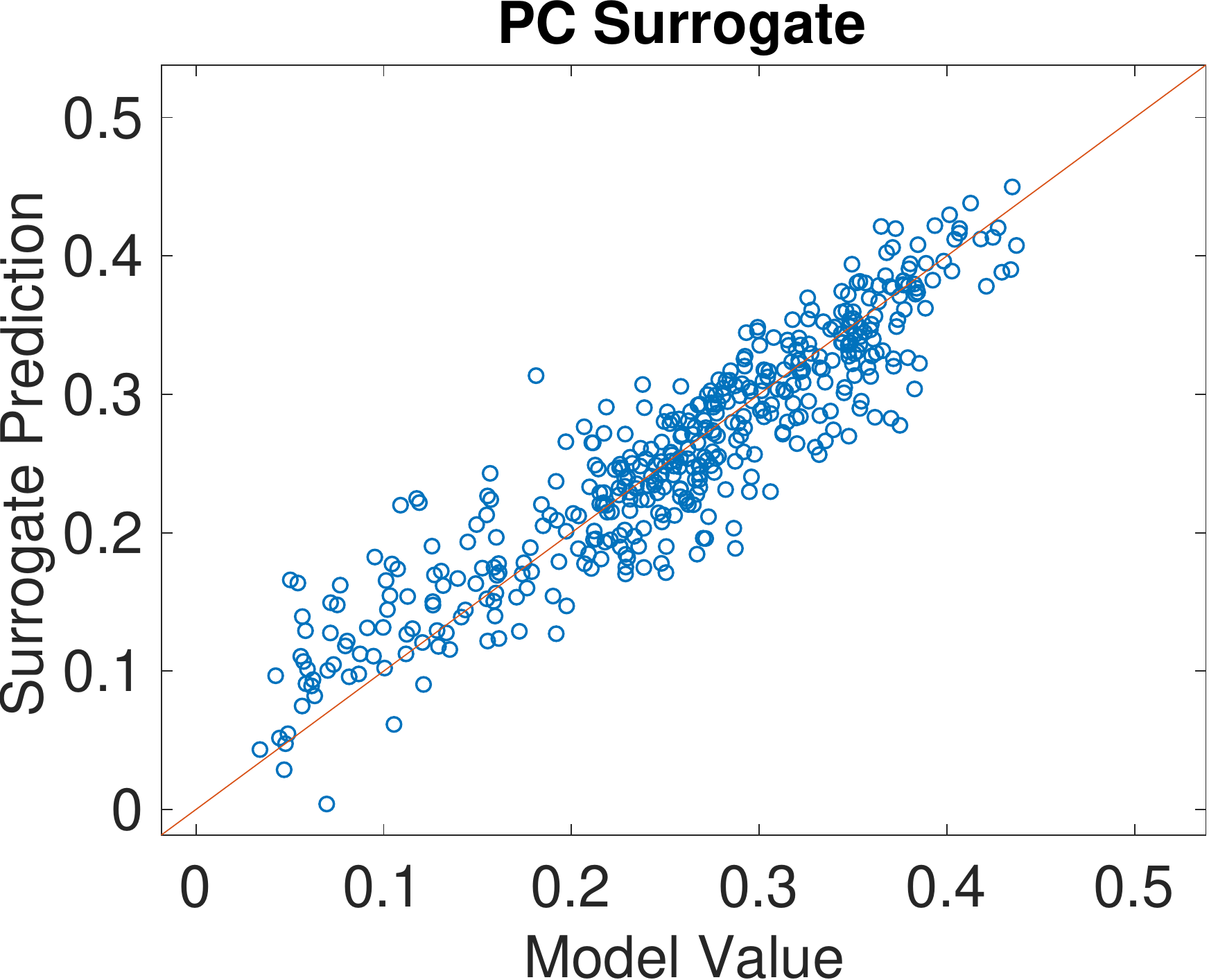}
\hspace{.1 cm}
\includegraphics[width=.475 \textwidth]{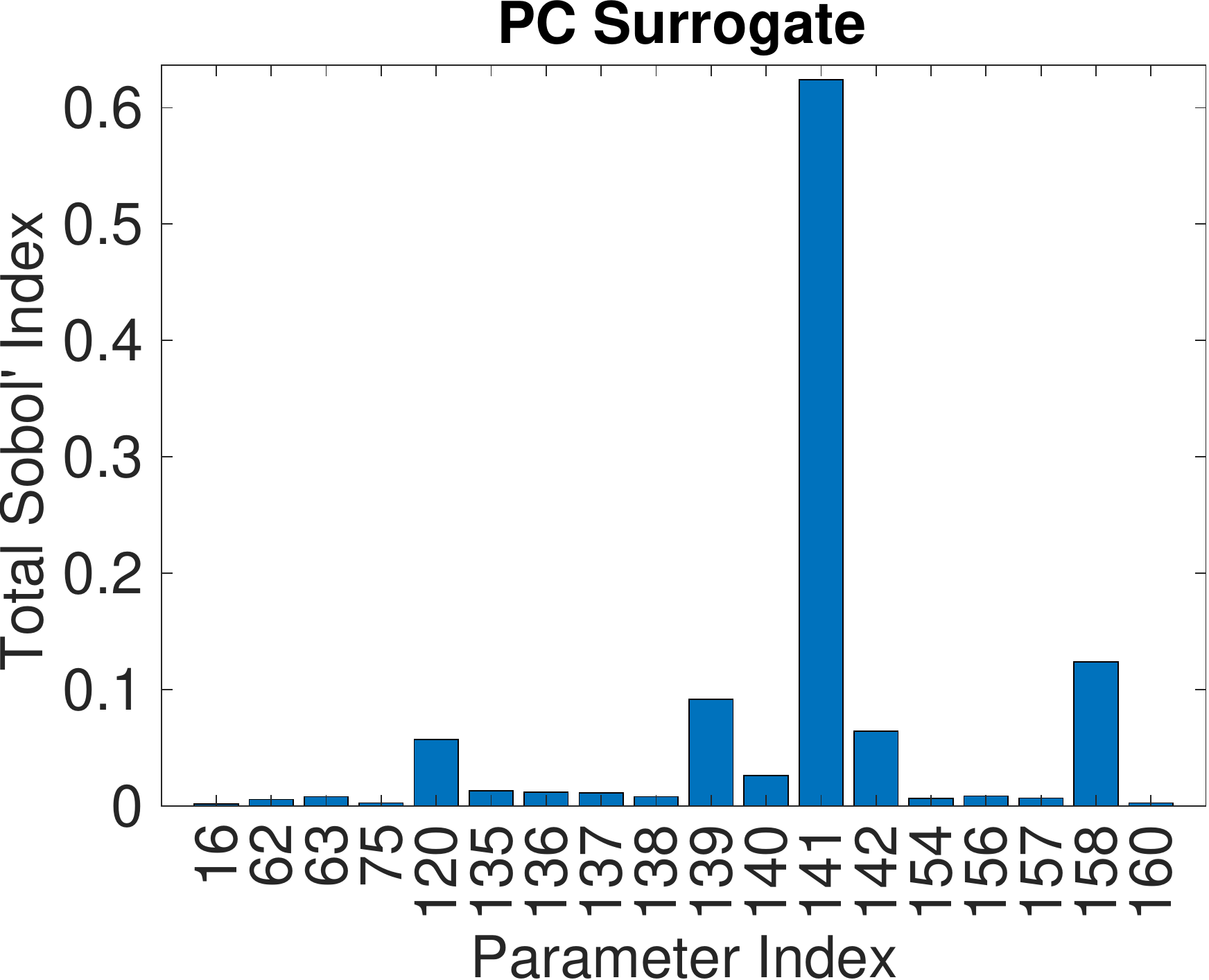}
\caption{$[AM+AM_p]_{min}$ QoI with experimental pulse stimulus. From left to right and top to bottom: linear surrogate predictions, linear surrogate importance measure, PC surrogate predictions, total Sobol' indices for PC surrogate.}
\label{fig:qoi_AM_AMp_Min_exp}
\end{figure}

Figure~\ref{fig:qoi_AM_AMp_Min_exp} displays results for the minimum of the combined concentration of the actin myosin complex with the experimental pulse stimulus. Table~\ref{tab:qoi_AM_AMp_Min} reports the five most important parameters and their total Sobol' indices. By comparing Table~\ref{tab:qoi_AM_AMp_Min} and Table~\ref{tab:qoi_vol_flow}, we see that the $[AM+AM_p]_{min}$ and average volumetric flow rate QoIs share four of their five most important parameters. Although at first analysis this should not be surprising it does suggest that the reaction rates of the actin myosin model for vessel contraction/dilation are relatively insensitive to finding the minimum of the contraction force and that the KIR ion channel is a vital component of the model. 

The linear surrogate does not perform as well for this QoI; however, the PC surrogate is far more accurate; this highlights the nonlinearity of this particular QoI. 

 As in the previous results, parameter $\theta_{120}$ appears important in the linear surrogate and less important in the PC surrogate, albeit, it is more important for this QoI than the previous ones.

\begin{table}[h]
\centering
\ra{1.3}
\begin{tabular}{cccc}
\toprule
Parameter & Identification in Supplementary Material & Total Sobol' Index (exp.) & Total Sobol' Index (rect.) \\
\midrule
$\theta{141}$ & $z_4$ in equation (149)  & 0.6242 & 0.6203\\
$\theta{158}$ & $n_{cross}$ in equation (214)  & 0.1239 & 0.1488\\
$\theta{139}$ &  $z_2$ in equation (148)  & 0.0918 & 0.0954\\
$\theta{142}$ & $z_5$ in equation (149)  &0.0644 & 0.0629\\
$\theta{120}$ & Nominal value 5.5 in equation. (10)  & 0.0571 & 0.0240\\
 \arrayrulecolor{black}\bottomrule
\end{tabular}
\caption{Five most influential parameters for the $[AM+AM_p]_{min}$ QoI when the experimental pulse stimulus is applied. The leftmost column is the parameter, the left-center column identifies the parameter in the Supplementary Material, the right-center column is the total Sobol' index computed for the parameter using the experimental pulse stimulus, and the right column is the total Sobol' index computed for the parameter using the rectangular pulse stimulus.}
\label{tab:qoi_AM_AMp_Min}
\end{table}

\section{Discussion}
We  note at the outset that the sensitivity analysis investigates the numerical model rather than the physiological one. However, since the numerical model has been validated with experimental data  \cite{Dormanns2015,Mathias2018},  the results indicate parameters (and therefore areas) of importance for physiologists and modellers. The integration times $t_1,t_2$ are the start and end of neuronal stimulation. Decreasing or increasing these time does not significantly alter the ranking of the parameters. We discuss below results for each QoI individually.

\subsection{Average ECS Potassium}
It is interesting to note for this particular quantity of interest that the first and third most important parameters are associated not with a \pot channel but the persistent \na channel, NaP.  The neuron model used in this analysis was developed from the work of Kager et al \cite{Kager2000a} and Chang et al \cite{Chang2013}.
The differential equation governing $m_4$,  the activation variable for the NaP channel, is given by equation (\ref{eqn:m4}).

\begin{figure}
\centering
\includegraphics[width=0.49\linewidth]{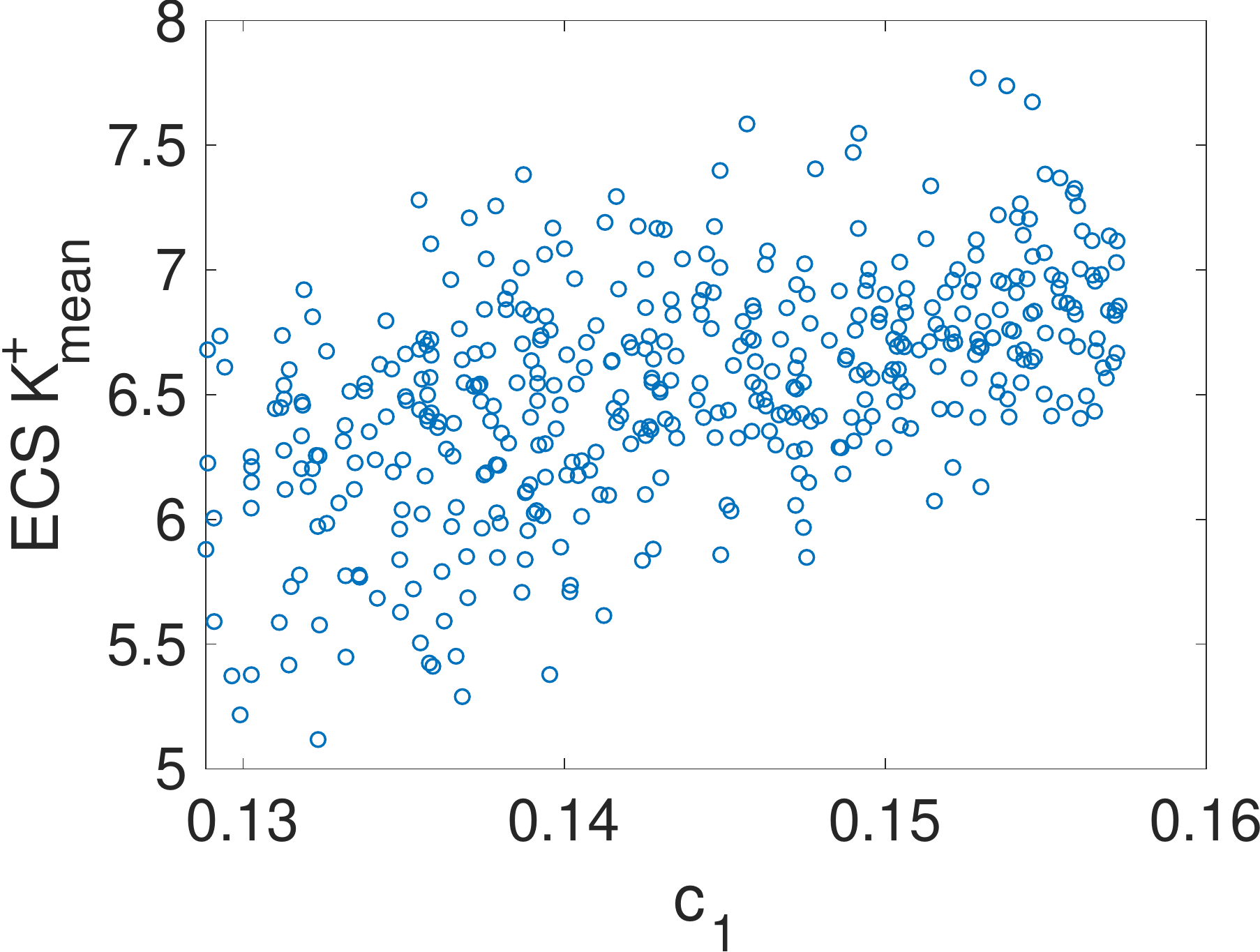}
\includegraphics[width=0.48\linewidth]{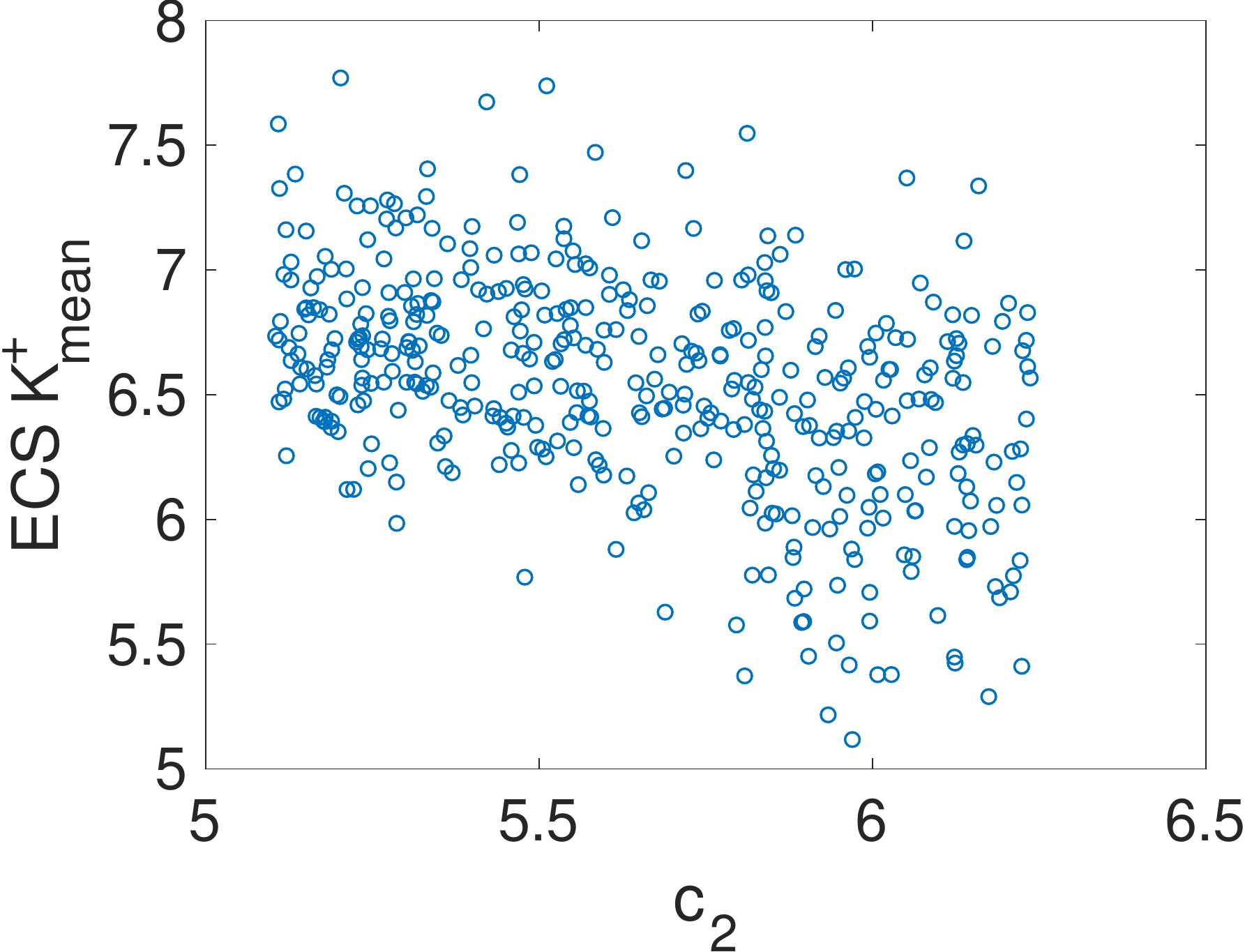}
\caption{Left: scatter plot of $\theta_{62}$ and the average ECS potassium \eqref{K_ECS_Mean}; right: scatter plot of $\theta_{63}$ and the average ECS potassium \eqref{K_ECS_Mean}.}
\label{fig:scatter}
\end{figure}

The inactivation of the NaP channel is very small compared to activation and therefore these parameters make little impact on the extracellular \pot. 
Scatter plots for the average of \pot in the  ECS against the parameters $\theta_{61}$ and $\theta_{62}$ are shown in Figure \ref{fig:scatter}.  For each parameter, there is a linear trend where increasing the parameter yields either an increase or decrease in the \pot ECS. The characteristic time for the activation variable for the NaP channel is defined as $\tau=\frac{1}{m_{4 \alpha}+m_{4 \beta}}$ which, by using equation (\ref{eqn:m4}), is constant (6 ms). The scatter plots also indicate this definition. These results show that a variation of $\pm 10 \%$ in either $\theta_{62}$ or $\theta_{63}$ can either reduce or increase the extracellular \pot  by approximately $15 \%$. We should note that these results do not take into account the spatial buffering carried out by the astrocytic syncytium which may have a significantly greater effect on the extracellular \pot \cite{Bellot-Saez2017,Kenny2018b}. 

The main driver for \pot in the extracellular space is the \pot \na ATP-ase pump which has the general form given by (in the dendrite)  
\begin{eqnarray}\label{eqn:ATP-ase}
\frac{(Na^{+}_{d})^3}{\left( Na^{+}_{d}+Na^{+}_{d,baseline}\right)^3}\frac{(K^{+}_e)^{2}}{\left( K^+_e +K^+_{e,baseline}\right) ^2}.
\end{eqnarray}
The ATP-ase pumps out \na and in \pot in the ratio of 3 to 2, hence the \na in the dentrite has a large effect on the extracellular \pot as seen in equation (\ref{eqn:ATP-ase}) and strengthens the result that the NaP channel has the most effect on the \pot. One could have expected the NaT (transient \na) channel to be prominent however, it has a fast inactivation variable and, although it produces a larger flux, does so over a shorter time.  
\subsection{Average Volumetric Flow Rate}
In the presented model, the main pathway for neurovascular coupling is that of the \pot pathway. Here the astrocyte takes up \pot ~from the synaptic cleft and provides an efflux into the perivascular space (PVS) via the BK channel. The smooth muscle cell detects this increase in PVS \pot and through the inwardly rectifying channel, $K_{IR}$, hypopolarises the SMC, shutting off the voltage mediated \ca channel.  \ca is therefore reduced and the SMC dilates. As stated in the Section~\ref{sec:results}, the most important parameter is the shift parameter $\theta_{141}$ in the $K_{IR}$ conductance \eqref{eqn:gkir} determining the magnitude of ion flux per unit change in the membrane potential away from the equilibrium (Nernst potential).  Compared to the other parameters defining the $K_{IR}$ conductance, $\theta_{141}$ is large with a nominal value of 12.6 whereas $\theta_{140}=4.2 \times 10^{-4} \mu M^{-1}$ and $\theta_{142} = -7.4 \times10^{-2} mV^{-1}$. Hence, for constant membrane potential and \pot,   variations in $\theta_{141}$ produce exponentially large variations in the $K_{IR}$ channel conductance allowing substantial efflux of \ca from the SMC and a dilation of the vessel $\left(\frac{dR}{dt}> 0\right)$. \\

The second most important parameter is the power index for the cytosolic SMC \ca which mediates the four state latch model of Hai and Murphy \cite{Hai1988} that is used in this model, in particular the rate of phosphorylation of  myosin and the actin-myosin complex.  Variations in \ca as predominantly dictated by the $K_{IR}$ channel will therefore have a direct effect on the dilation/contraction properties of the SMC and hence the perfusing vessel radius.  
The remaining three parameters have significantly lower total Sobol' indices and therefore only make small contributions to the QoI.
 
\subsection{$[AM+AM_p]_{min}$}
We see similar parameters appearing as most important for this QoI and the average volumetric flow rate QoI, which is to be expected given the strong relationship between the radius dilation/contraction phenomenon and the total quantity of actin/myosin complex and its phosphorylated compliment. In fact, this QoI serves as a test of the statistical mechanism in that the only two non-repeating parameters (in the list of five most important) between the volumetric flow rate QoI and the actin/myson complex QoI are $\theta_{120}$ and $\theta_{136}$, which, as noted above, are significantly less import than the leading parameters.
\section{Conclusion}
 A three stage methodology is presented for global sensitivity analysis of numerical cell models with a large number of parameters. To the authors knowledge, this is the first type of analysis which investigates neuroscience models of such size. The analysis investigated three quantities of interest pertaining to a numerical model of neurovascular coupling. The results indicated several prevalent features of the model. A significant influence of the persistent $Na^+$ channel activation variable on the average extracellular space $K^+$, a two parameter set (the inwardly rectifying $K^+$ channel shift parameter and the index for the cytosolic $Ca^{2+}$) which characterizes most of the variability in the average volumetric flow rate, and strong similarities between the most influential parameters for the average volumetric flow rate and minimum value of the combined actin/myocin complex.
 
In addition to the results reported in this article, four other QoI's were considered. Two of them, the maximum and average potassium concentration in the Astrocyte, were omitted because the parameter to QoI mapping is nearly constant and hence global sensitivity analysis is not necessary. Specifically, the mean of the QoI is approximately 37 times larger than its standard deviation for both cases. The other two unreported QoIs correspond to lag times. The first being the duration of time between the application of the stimulus and the minimum of the phosphorylated actin myosin complex, and the second being the duration of time between the application of the stimulus and the maximum of the radius. In both cases, the QoI exhibited highly nonlinear behavior which we were unable to approximate with linear or PC surrogates trained on the existing data. In fact, fitting such nonlinearities would likely require more samples than is computationally feasible for this model. The linear surrogate had 59\% and 47\% relative $L^2$ errors for these two QoIs, respectively. Our global sensitivity analysis methodology was unsuccessful because the linear surrogate was an unreliable tool for screening. Defining the QoI as the maximum/minimum value instead of the time lag makes the analysis more tractable. These maximum/minimum value QoIs were also considered and yielded similar results to the QoIs reported in the article.

For a given model and collection of samples, the methodology presented in this article may be applicable for some QoIs and intractable for others. The success of our method depends upon the surrogate models being sufficiently accurate. A general principle is that QoIs defined as averages will be more amenable for analysis than, for instance, minimum values or lag times. A practical benefit of our method is that any QoI may be considered without requiring additional model evaluations. The sampling and ODE solves are executed once, followed by computing the QoIs and performing global sensitivity analysis, which may be easily repeated for many different QoIs.
 
\section*{Appendix}

Determining the coefficients $c_{\boldsymbol{\alpha}}$ in the Polynomial Chaos surrogate \eqref{pce} is challenging. Ideally, one would solve the least squares problem
\begin{ceqn}
\begin{eqnarray}
\label{least_squares}
\min \sum\limits_{k=1}^M \left(g(\boldsymbol{\theta}^k)-\sum_{\boldsymbol{\alpha}} c_{\boldsymbol{\alpha}} \psi_{\boldsymbol{\alpha}}(\hat{\boldsymbol{\theta}}) \right)^2 \tag{A.1}
\end{eqnarray}
\end{ceqn}
to determine the coefficients. This approach is not currently feasible for the problems considered in this article. If there are, for instance, 18 input variables ($\theta_{j_i}$'s) then a $3^{rd}$ degree polynomial has 1330 unknown coefficients and a $4^{th}$ degree polynomial has 7315 unknown coefficients. With less than 1000 sample points, as in our case, \eqref{least_squares} will admit infinitely many solutions which interpolate the data but will yield poor approximations of the QoI. Rather, we seek an approximate solution of \eqref{least_squares} for which most of the coefficients are exactly 0. This may be achieved by introducing a penalty term and solving
\begin{ceqn}
\begin{eqnarray}
\label{least_squares_reg}
\min \sum\limits_{k=1}^M \left(g(\boldsymbol{\theta}^k)-\sum_{\boldsymbol{\alpha}} c_{\boldsymbol{\alpha}} \psi_{\boldsymbol{\alpha}}(\hat{\boldsymbol{\theta}}) \right)^2 + \lambda \sum_{\boldsymbol{\alpha}} \vert c_{\boldsymbol{\alpha}} \vert \tag{A.2}
\end{eqnarray}
\end{ceqn}
instead of \eqref{least_squares}. Adding the sum of absolute values of the coefficients encourages a sparse solution, i.e. one with many 0 coefficients. However, it comes at the cost of making the objective function non-differentiable and hence \eqref{least_squares_reg} requires a more sophisticated optimization approach in comparison to \eqref{least_squares}. A plurality of well documented methods exist for solving \eqref{least_squares_reg}. In this article we use Least Angle Regression (LAR) \cite{lar} with its implementation in \cite{uqlab}, and a maximum polynomial degree of 5.

Because the basis function of the Polynomial Chaos surrogate are orthogonal with respect to the PDF $p_{\hat{\boldsymbol{\theta}}}$, the variance and conditional expectation in \eqref{sobol} may be computed analytically as a function of the coefficients. Hence the total Sobol' indices of the Polynomial Chaos surrogate are given in closed form as a function of the coefficients.

\bibliographystyle{spbasic} 
\bibliography{library}

\end{document}